\documentclass[aps,prd,twocolumn,preprintnumbers,showpacs,nofootinbib,amssymb]{revtex4}
\usepackage{graphicx}
\usepackage{amsmath}
\usepackage{amssymb}
\usepackage{amsfonts}
\usepackage{bm}
\usepackage{color}

\def\be{\begin{equation}}
\def\ee{\end{equation}}
\def\bea{\begin{eqnarray}}
\def\eea{\end{eqnarray}}

\begin{document}

\title{Cosmology with a stiff matter era}
\author{Pierre-Henri Chavanis}
%\email{chavanis@irsamc.ups-tlse.fr}
\affiliation{Laboratoire de Physique Th\'eorique,
Universit\'e Paul Sabatier, 118 route de Narbonne  31062 Toulouse, France}
%\author{Tiberiu Harko}
%\email{harko@hkucc.hku.hk}
%\affiliation{Department of Physics and
%Center for Theoretical and Computational Physics, The University
%of Hong Kong, Pok Fu Lam Road, Hong Kong, P. R. China}

\begin{abstract}
We provide a simple analytical solution of the Friedmann equations for a
universe made of stiff matter, dust matter, and dark energy. A stiff matter era
is present in the cosmological model of Zel'dovich (1972) where the primordial
universe is assumed to be made of a cold gas of baryons. It also occurs
in certain cosmological models where dark matter is made of relativistic
self-gravitating Bose-Einstein condensates (BECs). When the energy
density of the stiff matter is positive, the primordial universe is singular.
It starts from a state with a vanishing scale factor and an infinite density. We
consider the possibility that the energy density of the stiff matter is negative
(anti-stiff matter). This happens, for example, when the BECs have an attractive
self-interaction.
In that case, the primordial universe is non-singular. It starts from a state
in which the scale factor is finite and the energy density is equal to zero. For
the
sake of generality, we consider a cosmological constant of arbitrary sign. When
the cosmological constant is positive, the universe asymptotically reaches a de
Sitter phase where the scale factor increases exponentially rapidly. This can
account for the accelerating expansion of the universe that we observe at
present. When the cosmological constant is negative (anti-de Sitter), the
evolution of the universe is cyclic. Therefore, depending on the sign of the
energy density of the stiff matter and of the dark energy, we obtain singular
and non-singular expanding or cyclic universes. 
\end{abstract}


\maketitle

\section{Introduction}
\label{sec_intro}

In a seminal paper, Zel'dovich \cite{zeldocosmo} introduced a cosmological model
in which the very early universe is assumed to be made of a cold gas of baryons
with a stiff equation of state $P=\epsilon$, where $P$ is the pressure and
$\epsilon$ is the energy density. For  a stiff matter fluid, the velocity of
sound $c_s=\sqrt{P'(\epsilon)}c$ is equal to the velocity of light
\cite{zeldovich}. In that
case, the energy density decreases as $\epsilon\propto 1/a^6$ where $a(t)$ is
the scale factor.
This phase, if it has ever existed, preceded the radiation era ($P=\epsilon/3$,
$\epsilon\propto a^{-4}$), the dust matter era ($P=0$, $\epsilon\propto
a^{-3}$), and the dark energy era ($P=-\epsilon$,
$\epsilon=\epsilon_{\Lambda}$). A stiff matter era also occurs in certain
cosmological models
in which dark matter is made of relativistic self-gravitating BECs
\cite{mlbec}. In that case, the energy density of the stiff
matter can be
positive or negative depending whether the self-interaction of the bosons is
repulsive or attractive. 

We show that when radiation is neglected, the Friedmann equations for a
universe made of stiff matter, dust matter, and dark energy can be
integrated analytically. This provides a simple cosmological model exhibiting
a stiff matter era. We consider the general case where the energy
density of the stiff fluid is positive or negative. The case of a stiff fluid
with a  positive energy density is very similar to the standard
model of cosmology in the
sense that the universe begins by a singularity at $t=0$ in which the scale
factor is
equal to zero while the energy density is infinite. Initially,
the scale factor increases as $a(t)\propto t^{1/3}$
and the energy density decreases as $\epsilon(t)\propto t^{-2}$. Interestingly,
the presence of a stiff fluid with a negative energy density (anti-stiff fluid)
prevents the primordial singularity. In that case, we obtain a model of
universe in which the initial scale factor is finite and the energy density is
equal to zero. For the sake of
generality, we consider a positive or a
negative cosmological constant. At late times, a dark fluid with a
positive energy density leads to a de Sitter phase in which the scale
factor increases exponentially rapidly. This can account for the present
acceleration of the universe. By contrast, when the dark fluid has a negative
energy density (anti-dark fluid) the evolution of the universe is cyclic. 

The paper is organized as follows. In Sec. \ref{sec_stiffeos}, we consider a
perfect fluid at
$T=0$ described by a polytropic equation of state of the form
$P=K\rho^2$ where $\rho$ is rest-mass density. We determine the relation between
the energy density $\epsilon$ and
the rest-mass density $\rho$ and obtain an
explicit equation of state $P(\epsilon)$ relating the pressure to the energy
density. At high energies, it reduces to a stiff equation of state $P\simeq
\epsilon$ for which the velocity of sound is equal to the velocity of light.
This equation of state describes a cold gas of baryons in Zel'dovich model
\cite{zeldocosmo} or
a relativistic gas of BECs (in certain approximations) \cite{mlbec}. In Secs.
\ref{sec_cosmostiff} and \ref{sec_stiff},  we apply this equation of state to
cosmology. This leads to a cosmological model exhibiting a primordial stiff
matter era, followed by a radiation era, a dust matter era, and a dark energy
era. In Secs. \ref{sec_pp}-\ref{sec_rad}, we provide simple analytical solutions
of the Friedmann
equations for a universe comprising a stiff fluid. We consider four cases: (i)
a singular expanding universe
with $K\ge 0$ and $\Lambda\ge 0$; (ii)  a non-singular expanding universe
with $K\le 0$ and $\Lambda\ge 0$; (iii) a singular cyclic universe
with $K\ge 0$ and $\Lambda\le 0$; (iv) a non-singular cyclic universe
with $K\le 0$ and $\Lambda\le 0$.

\section{A stiff equation of state}
\label{sec_stiffeos}

In this section, we consider physical systems described by a stiff equation of
state.

\subsection{General results for a fluid at $T=0$}
\label{sec_gr}

The local form of the first law of thermodynamics can be written as
\begin{equation}
d\left (\frac{\epsilon}{\rho}\right )=-P d\left (\frac{1}{\rho}\right )+T d\left
(\frac{s}{\rho}\right ),
\label{gr1}
\end{equation}
where $\rho$ is the mass density and $s$ is the
entropy density in the rest frame. For a system at $T=0$, the first law of
thermodynamics reduces to
\begin{equation}
d\epsilon=\frac{P+\epsilon}{\rho}d\rho.
\label{gr2}
\end{equation}
For a given equation of state, Eq. (\ref{gr2}) can be integrated to obtain the
relation between the energy density $\epsilon$ and the rest-mass density $\rho$.
For example, let us assume that the equation of state is prescribed under the
form $P=P(\rho)$. In that case, Eq. (\ref{gr2}) reduces to the first order
linear differential equation
\begin{equation}
\frac{d\epsilon}{d\rho}-\frac{1}{\rho}\epsilon=\frac{P(\rho)}{\rho}.
\label{gr3}
\end{equation}
Using the method of the variation of the constant, we obtain
\begin{equation}
\epsilon=A\rho c^2+\rho\int^{\rho}\frac{P(\rho')}{{\rho'}^2}\, d\rho',
\label{gr4}
\end{equation}
where $A$ is a constant of integration. For an equation of state
$P(\rho)$ such that $P\sim \rho^{\gamma}$ with
$\gamma>1$ when $\rho\rightarrow 0$, we  determine the constant $A$ in Eq.
(\ref{gr4}) by
requiring that $\epsilon\sim\rho c^2$ when $\rho\rightarrow 0$. This gives
\begin{equation}
\epsilon=\rho c^2+\rho\int_0^{\rho}\frac{P(\rho')}{{\rho'}^2}\, d\rho'=\rho
c^2+u(\rho).
\label{gr5}
\end{equation}
We note that $u(\rho)$ may be interpreted as an internal energy
\cite{aaantonov}.

\subsection{Polytrope $n=1$}
\label{sec_poly}

We consider the equation of state
\begin{equation}
P=K\rho^{2},
\label{poly1}
\end{equation}
corresponding to a polytrope of index $n=1$ \cite{chandra}. In that case, Eq.
(\ref{gr5}) reduces to
\begin{equation}
\epsilon=\rho c^2+P=\rho c^2+{K}\rho^{2}.
\label{poly2}
\end{equation}
This equation can be reversed to give
\begin{equation}
\rho=\frac{c^2}{2K}\left (\sqrt{1+\frac{4K\epsilon}{c^4}}-1\right ).
\label{poly3}
\end{equation}
Combining Eqs. (\ref{poly1}) and (\ref{poly3}), we obtain the relation
between the pressure and the energy density
\begin{equation}
P=\frac{c^4}{4K}\left (\sqrt{1+\frac{4K\epsilon}{c^4}}-1\right )^2.
\label{poly4}
\end{equation}
For $\epsilon\rightarrow 0$ (non-relativistic limit), we get
\begin{eqnarray}
\epsilon\sim \rho c^2,\qquad P\sim \frac{K}{c^4}\epsilon^2,\qquad P= K\rho^2.
\label{poly5}
\end{eqnarray}
This is a polytropic equation of state $P=K(\epsilon/c^2)^2$ of
index $n=1$. For $\epsilon\rightarrow +\infty$ (ultra-relativistic limit), we
get
\begin{eqnarray}
\epsilon\sim K\rho^{2},\qquad P\sim \epsilon,\qquad P= K\rho^{2}.
\label{poly6}
\end{eqnarray}
This is a linear equation of state $P=\epsilon$ that is called ``stiff''
because the velocity of sound is equal to the velocity of light ($c_s=c$).

For the equation of state (\ref{poly4}), the velocity of sound is given by
\begin{equation}
\label{poly7}
\frac{c_s^2}{c^2}=P'(\epsilon)=1-\frac{1}{\sqrt{1+4K\epsilon/c^4}}.
\end{equation}
We always have $c_s<c$. For $\epsilon\rightarrow +\infty$,  $c_s\rightarrow c$.

\subsection{Gas of baryons interacting through a vector meson field}
\label{sec_baryon}

Zel'dovich \cite{zeldocosmo,zeldovich} introduced a cosmological model
in which the primordial universe is made of a gas of baryons interacting through
a vector meson field and showed that the equation of state of this system is of
the form of Eq. (\ref{poly1}) with a polytropic constant
\begin{equation}
\label{baryon1}
K=\frac{g^2 h^2}{2\pi m_m^2 m_b^2 c^2},
\end{equation}
where $g$ is the baryon charge, $m_m$ is the meson mass, and $m_b$ is the baryon
mass. Zel'dovich \cite{zeldovich} introduced this equation of state as an
example to show how the speed of sound could approach the speed of light at very
high pressures and densities.

Zel'dovich \cite{zeldocosmo,zeldovich} also mentions that
the complete equation of state of his model is of the form
\begin{eqnarray}
P=K\rho^2+K'\rho^{4/3},
\label{baryon2}
\end{eqnarray}
where $K$ is given by Eq. (\ref{baryon1}) and the second term accounts for
quantum (Fermi) corrections. For the equation of state (\ref{baryon2}), we find
from Eq. (\ref{gr5}) that the relation between the energy density and the
rest-mass density is
\begin{eqnarray}
\epsilon=\rho c^2+K\rho^2+3K'\rho^{4/3}.
\label{baryon3}
\end{eqnarray}

\subsection{Relativistic self-gravitating BECs}
\label{sec_gpp}

Some authors have proposed that dark matter may be made of self-gravitating BECs
with short-range
interactions
\cite{leekoh,peebles,goodman,arbey,lesgourgues,bohmer,briscese,harko,pires,
rmbec,rindler,lora,lensing,glgr1,msepl,prd1,prd2}. In the TF approximation, a
BEC is
equivalent to a fluid with
an equation of state of the form of Eq. (\ref{poly1}) with a polytropic
constant
\begin{equation}
\label{gpp1}
K=\frac{2\pi \hbar ^{2}a_s}{m^{3}},
\end{equation}
where $m$ is the mass of the bosons and $a_s$ is their scattering length.
This equation of state can be derived from the classical Gross-Pitaevskii
equation \cite{gross,pitaevskii} after writing it under the form of fluid
equation by using the Madelung transformation \cite{madelung}. It is therefore a
non-relativistic equation of state that, in principle, is not valid in the
relativistic regime. Nevertheless, we can consider a partially-relativistic
model in which we use the classical equation of state (\ref{poly1}) with  the
relativistic relation (\ref{poly2}) between the energy density and the rest-mass
density. This approximation has been considered in \cite{mlbec}.

A feature of BECs is that the scattering length $a_s$ can be positive or
negative \cite{revuebec}. Positive values of $a_s$ correspond to repulsive
interactions and
negative values of $a_s$ correspond to attractive interactions. When $a_s$ is
positive, the pressure is positive and when $a_s$ is negative the pressure is
negative. We shall consider the two possibilities in the following.

\section{Cosmology with a stiff fluid}
\label{sec_cosmostiff}

In this section, we derive the main equations governing the evolution of a
universe described by a stiff equation of state of the form of Eq.
(\ref{poly4}).

\subsection{The Friedmann equations}
\label{sec_friedmann}

We assume that the universe is homogeneous and isotropic, and contains a uniform
perfect fluid of energy density $\epsilon(t)$ and isotropic pressure $P(t)$. The
radius of curvature of the $3$-dimensional space, or scale factor, is noted
$a(t)$ and the curvature of space is noted $k$. The universe is closed if $k>0$,
flat if $k=0$, and open if $k<0$. We assume that the universe is flat ($k=0$) in
agreement with the observations of the cosmic microwave background (CMB)
\cite{bt}. In that case, the Einstein equations can be written as
\cite{weinbergbook}:
\begin{equation}
\frac{d\epsilon}{dt}+3\frac{\dot a}{a}(\epsilon+P)=0,
\label{f1}
\end{equation}
\begin{equation}
\frac{\ddot a}{a}=-\frac{4\pi G}{3c^2}\left(\epsilon+3P\right
)+\frac{\Lambda}{3},
\label{f2}
\end{equation}
\begin{equation}
H^2=\left (\frac{\dot a}{a}\right )^2=\frac{8\pi
G}{3c^2}\epsilon+\frac{\Lambda}{3},
\label{f3}
\end{equation}
where we have introduced the Hubble parameter $H=\dot a/a$ and accounted for a
possible non-zero cosmological constant $\Lambda$. The cosmological constant is
equivalent to a dark energy fluid with a constant density
\begin{equation}
\epsilon_{\Lambda}=\rho_{\Lambda}c^2=\frac{\Lambda c^2}{8\pi G},
\label{f4}
\end{equation}
and an equation of state $P=-\epsilon$. Eqs. (\ref{f1})-(\ref{f3}) are the
well-known Friedmann equations describing a
non-static universe. Among these three equations, only two are independent. The
first equation can be viewed as an equation of continuity. For a given
barotropic equation of state $P=P(\epsilon)$, it determines the relation between
the energy density $\epsilon$ and the scale factor $a$. Then, the evolution of
the scale factor $a(t)$ is given by Eq. (\ref{f3}).

\subsection{General results for a fluid at $T=0$}
\label{sec_gent}

We assume that the universe is made of a fluid at  $T=0$ with an equation of
state $P(\rho)$. In that case, the relation between the energy density
$\epsilon$ and the rest-mass density $\rho$ is given by the first law of
relativistic
thermodynamics, Eq. (\ref{gr2}).  Combining this relation with the continuity
equation (\ref{f1}), we get
\begin{equation}
\frac{d\rho}{dt}+3\frac{\dot a}{a}\rho=0.
\label{eos2}
\end{equation}
We note that this equation is exact for a fluid at $T=0$ and that it does not
depend on the explicit form of the equation of state $P(\rho)$. It can be
integrated into
\begin{equation}
\rho=\rho_0 \left (\frac{a_0}{a}\right )^3,
\label{eos3}
\end{equation}
where $\rho_0$ is the present value of the rest-mass density and $a_0$ is the
present
value of the scale factor.

\subsection{Polytrope $n=1$}
\label{sec_polyb}

We now consider the equation of state (\ref{poly1}). As we have seen, this
equation of state appears in the cosmological model of Zel'dovich
\cite{zeldocosmo} in which the early universe is made of a cold
gas of baryons. This equation
of state also appears in certain cosmological models in which dark matter is
made of
relativistic BECs \cite{mlbec}. For the equation of state (\ref{poly1}), Eq.
(\ref{gr5})
can be integrated easily and the relation between the energy density and the
rest-mass density is given by Eq. (\ref{poly2}). Combining Eqs. (\ref{poly2})
and (\ref{eos3}), we get
\begin{equation}
\epsilon=\rho_0 c^2\left (\frac{a_0}{a}\right )^3+K\rho_0^2\left
(\frac{a_0}{a}\right )^6.
\label{eos6}
\end{equation}
This relation can also be obtained by solving the continuity equation (\ref{f1})
with the equation of state (\ref{poly4}), see Appendix D of \cite{mlbec}.

In the early universe ($a\rightarrow 0$), we have
\begin{equation}
\epsilon\sim K\rho_0^2\left (\frac{a_0}{a}\right )^6, \qquad \epsilon\sim
K\rho^2, \qquad P\sim\epsilon.
\label{eos7}
\end{equation}
These equations describe a stiff fluid ($P=\epsilon$) for which the velocity of
sound is equal to the velocity of light.

In the late universe ($a\rightarrow +\infty$), we have
\begin{equation}
\epsilon\sim \rho_0 c^2\left (\frac{a_0}{a}\right )^3, \qquad \epsilon\sim \rho
c^2, \qquad P\sim \frac{K}{c^4}\epsilon^2.
\label{eos8}
\end{equation}
These equations describe a fluid with a polytropic equation of
state  of index $n=1$ ($P=K\epsilon^2/c^4$). For very large values of
the
scale factor, we recover the results of the CDM model ($P=0$) since
$\epsilon\propto a^{-3}$.

When combined with the Friedmann equation  (\ref{f3}), Eq. (\ref{eos6})
describes a model of
universe exhibiting a stiff matter phase ($\epsilon\propto a^{-6}$),
a dust matter phase ($\epsilon\propto a^{-3}$), and a dark energy phase
($\epsilon\sim\rho_{\Lambda}c^2$).

\subsection{A more general equation of state}
\label{sec_moregen}

If we consider the more general equation of state (\ref{baryon2}) suggested by
Zel'dovich \cite{zeldovich,zeldocosmo}, the relation between the energy density
and the rest-mass density is given by
Eq. (\ref{baryon3}). Using Eq. (\ref{eos3}), we get
\begin{equation}
\epsilon=\rho_0 c^2\left (\frac{a_0}{a}\right )^3+K\rho_0^2\left
(\frac{a_0}{a}\right )^6+3K'\rho_0^{4/3}\left (\frac{a_0}{a}\right )^4.
\label{moregen1}
\end{equation}
When combined with the Friedmann equation  (\ref{f3}), Eq. (\ref{moregen1})
describes a model of
universe exhibiting a stiff matter phase ($\epsilon\propto a^{-6}$), a radiation
phase ($\epsilon\propto a^{-4}$), a dust matter phase ($\epsilon\propto
a^{-3}$), and a dark energy phase ($\epsilon\sim\rho_{\Lambda}c^2$).

\section{The Friedmann equations for a universe presenting a stiff matter era}
\label{sec_stiff}

In this section, we assume that the universe is made of one or several
fluids each of them described by a linear equation of state $P=\alpha\epsilon$.
The equation of continuity (\ref{f1}) implies that the energy density is related
to the scale factor by $\epsilon=\epsilon_0 (a_0/a)^{3(1+\alpha)}$, where the
subscript $0$ denotes present-day values of the quantities. A linear equation of
state can describe dust matter ($\alpha=0$, $\epsilon_m\propto a^{-3}$),
radiation ($\alpha=1/3$, $\epsilon_{rad}\propto a^{-4}$), stiff matter
($\alpha=1$, $\epsilon_s\propto a^{-6}$), vacuum energy ($\alpha=-1$,
$\epsilon=\epsilon_P$), and dark energy ($\alpha=-1$,
$\epsilon=\epsilon_{\Lambda}$).

More specifically, we consider a universe made of stiff matter, radiation, dust
matter and dark energy treated as non-interacting species. Summing the
contribution of each species, the total energy density can be written as
\begin{equation}
\epsilon=\frac{\epsilon_{s,0}}{(a/a_0)^6}+\frac{\epsilon_{rad,0}}{(a/a_0)^4}+\frac{\epsilon_{m,0}}{(a/a_0)^3}+\epsilon_{\Lambda}.
\label{stiff1}
\end{equation}
In this model, the stiff matter dominates in the early universe. This is
followed by the radiation era, by the dust matter era and, finally, by the dark
energy era. Writing $\epsilon_{\alpha,0}=\Omega_{\alpha,0}\epsilon_0$ for each
species, we get 
\begin{equation}
\frac{\epsilon}{\epsilon_0}=\frac{\Omega_{s,0}}{(a/a_0)^6}+\frac{\Omega_{rad,0}}
{(a/a_0)^4}
+\frac{\Omega_{m,0}}{(a/a_0)^3}+\Omega_{\Lambda,0}.
\label{stiff1b}
\end{equation}
The last term in Eq. (\ref{stiff1b}), corresponding to the dark energy, is
equivalent to the cosmological constant in Eq. (\ref{f3}). For the sake of
generality, we consider the case of a positive ($\Omega_{\Lambda,0}\ge 0$) or 
negative ($\Omega_{\Lambda,0}\le 0$)
cosmological constant $\Lambda$. When the radiation term is neglected, Eq.
(\ref{stiff1b}) is equivalent to Eq. (\ref{eos6}) obtained from the equation of
state (\ref{poly1}) and (\ref{baryon1}) proposed by Zel'dovich
\cite{zeldovich,zeldocosmo} or from the
equation of state  (\ref{poly1}) and (\ref{gpp1}) corresponding to a
relativistic BEC \cite{mlbec}.  On the other hand, Eq. (\ref{stiff1b}) with the
radiation
term included is equivalent to Eq. (\ref{moregen1}) obtained from the more
general equation of state (\ref{baryon2}) and (\ref{baryon1}) suggested by
Zel'dovich \cite{zeldovich,zeldocosmo}. Comparing Eqs. (\ref{eos6}) and
(\ref{stiff1b}), we see that a
positive value of $K$ corresponds to a positive energy density of the stiff
matter while a negative value of $K$ corresponds to
a negative energy density of the stiff matter. We shall therefore consider
the two possibilities $\Omega_{s,0}\ge 0$ and
$\Omega_{s,0}\le 0$. However, we take $\Omega_{m,0}\ge 0$ and
$\Omega_{rad,0}\ge 0$.

Using Eq. (\ref{stiff1b}), the Friedmann equation (\ref{f3}) takes the form
\begin{equation}
\frac{H}{H_0}=\sqrt{\frac{\Omega_{s,0}}{(a/a_0)^6}+\frac{\Omega_{rad,0}}{(a/a_0)^4}+\frac{\Omega_{m,0}}{(a/a_0)^3}+\Omega_{\Lambda,0}}
\label{stiff2}
\end{equation}
with $\Omega_{s,0}+\Omega_{rad,0}+\Omega_{m,0}+\Omega_{\Lambda,0}=1$ and
$H_0=(8\pi G\epsilon_0/3c^2)^{1/2}$.  We note the relation
\begin{equation}
\frac{\epsilon}{\epsilon_0}=\left
(\frac{H}{H_0}\right )^2. 
\label{stiff2b}
\end{equation}
The evolution of the
scale factor is given by
\begin{equation}
\int_{a_i/a_0}^{a/a_0} \frac{dx}{x \sqrt{\frac{\Omega_{s,0}}{x^6}+\frac{\Omega_{rad,0}}{x^4}+\frac{\Omega_{m,0}}{x^3}+\Omega_{\Lambda,0}}}=H_0 t,
\label{stiff3}
\end{equation}
where $a_i$ is the initial value of the scale factor determined below.

In Secs. \ref{sec_pp}-\ref{sec_nn},  we ignore radiation ($\Omega_{rad,0}=0$)
and consider a
universe made of stiff matter, dust matter, and dark energy. In that case, the
Friedmann equation (\ref{stiff3}) reduces to
\begin{equation}
\int_{a_i/a_0}^{a/a_0} \frac{dx}{x
\sqrt{\frac{\Omega_{s,0}}{x^6}+\frac{\Omega_{m,0}}{x^3}+\Omega_{\Lambda,0}}}=H_0
t,
\label{stiff3c}
\end{equation}
which can be integrated analytically. In Secs. \ref{sec_rad}, we provide some
particular analytical solutions of Eq. (\ref{stiff3}) in the
case where the
radiation is
taken into account.

\section{The case $\Omega_{s,0}\ge 0$ and $\Omega_{\Lambda,0}\ge 0$}
\label{sec_pp}

\begin{figure}[!ht]
\includegraphics[width=0.98\linewidth]{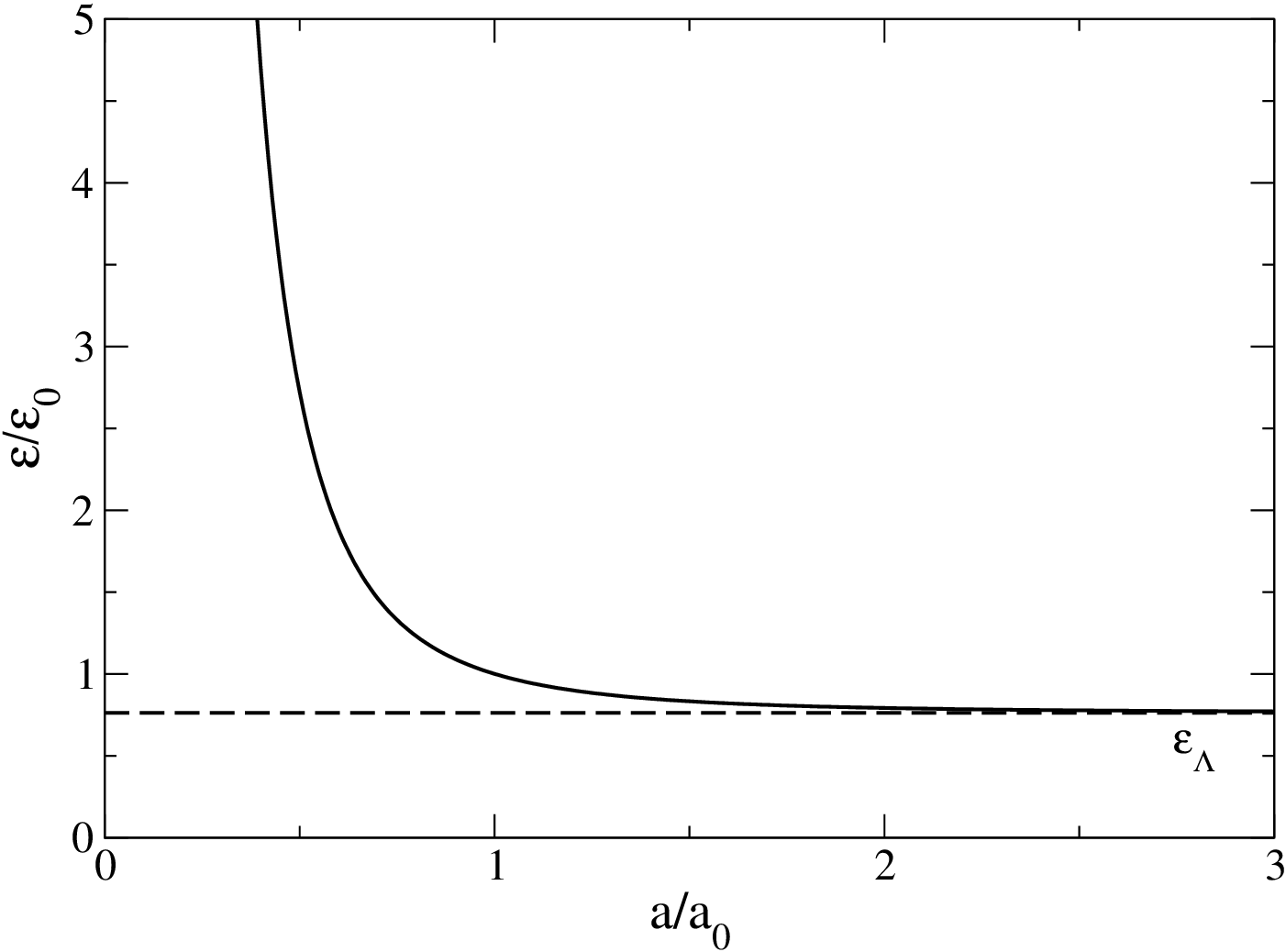}
\caption{Energy density as a function of the scale factor. We have taken
$\Omega_{m,0}=0.237$,
$\Omega_{\Lambda,0}=0.763$, and $\Omega_{s,0}=10^{-3}$ (here and in the
following figures, we have chosen a relatively large value of the density of
stiff matter $\Omega_{s,0}$ for a better illustration of the
results).\label{aepsPP}}
\end{figure}

\begin{figure}[!ht]
\includegraphics[width=0.98\linewidth]{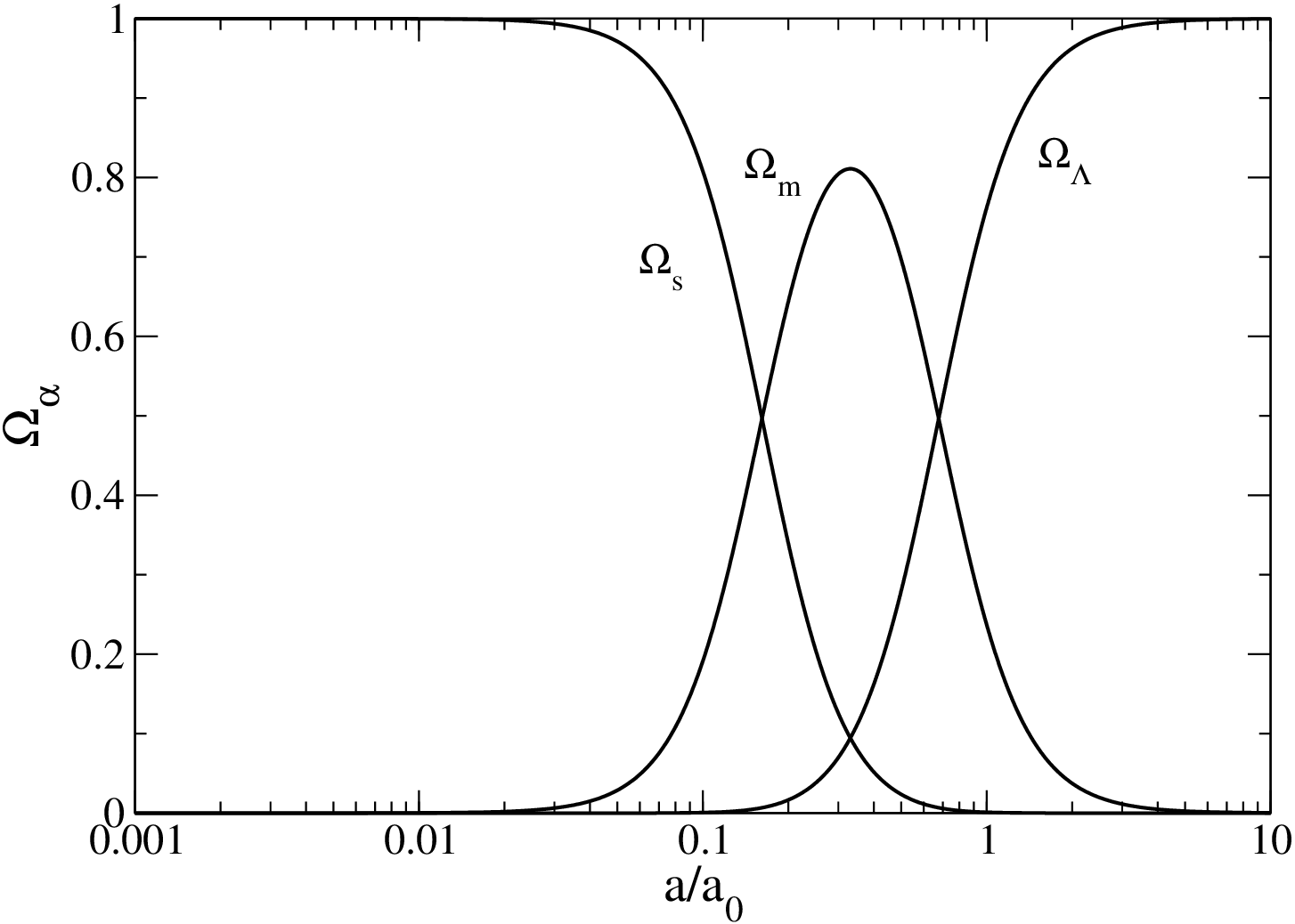}
\caption{Evolution of the proportion
$\Omega_{\alpha}=\epsilon_{\alpha}/\epsilon$ of the different components of the
universe with the scale factor. \label{proportionsPP}}
\end{figure}

We first consider the case of a positive stiff energy density ($\Omega_{s,0}\ge
0$) and a positive cosmological constant ($\Omega_{\Lambda,0}\ge 0$). The total
energy density is
\begin{equation}
\frac{\epsilon}{\epsilon_0}=\frac{\Omega_{s,0}}{(a/a_0)^6}+\frac{\Omega_{m,0}}{(a/a_0)^3}+\Omega_{\Lambda,0}.
\label{pp1}
\end{equation}
The energy density starts from $\epsilon=+\infty$ at $a=a_i=0$, decreases, and
tends to $\epsilon_{\Lambda}$ for $a\rightarrow
+\infty$. The relation between the energy density and the scale factor is shown
in Fig. \ref{aepsPP}. The proportions of stiff matter, dust matter and dark
energy as a function of the scale factor are shown in Fig.
\ref{proportionsPP}.

\subsection{Stiff matter, dust matter, and dark energy}
\label{sec_ppa}

We consider a universe made of stiff matter, dust matter, and dark energy. Using
the identity 
\begin{eqnarray}
\int \frac{dx}{x\sqrt{\frac{a}{x^3}+\frac{b}{x^6}+c}}\qquad\qquad\qquad\qquad\nonumber\\
=\frac{1}{3\sqrt{c}}\ln\left\lbrack a+2cx^3+2\sqrt{c}\sqrt{b+ax^3+cx^6}\right\rbrack,
\label{ppa1}
\end{eqnarray}
Eq. (\ref{stiff3c}) can be solved analytically to give 
\begin{eqnarray}
\frac{a}{a_0}=\Biggl \lbrack \left (\frac{\Omega_{m,0}}{\Omega_{\Lambda,0}}+2\sqrt{\frac{\Omega_{s,0}}{\Omega_{\Lambda,0}}}\right )\sinh^2\left (\frac{3}{2}\sqrt{\Omega_{\Lambda,0}}H_0 t\right )\nonumber\\
+\sqrt{\frac{\Omega_{s,0}}{\Omega_{\Lambda,0}}}\left (1-e^{-3\sqrt{\Omega_{\Lambda,0}}H_0 t}\right )\Biggr\rbrack^{1/3}.
\label{ppa2}
\end{eqnarray}
From Eq. (\ref{ppa2}), we can compute $H=\dot a/a$ leading to
\begin{eqnarray}
\left (\frac{a}{a_0}\right )^{3}\frac{H}{H_0}&=& \left
(\frac{\Omega_{m,0}}{2\sqrt{\Omega_{\Lambda,0}}}+\sqrt{\Omega_{s,0}}\right
)\sinh\left (3\sqrt{\Omega_{\Lambda,0}} H_0 t\right )\nonumber\\
&+&\sqrt{\Omega_{s,0}}
e^{-3\sqrt{\Omega_{\Lambda,0}}H_0 t}.
\label{ppa3}
\end{eqnarray}
The energy density
$\epsilon/\epsilon_0$ is given by Eq. (\ref{stiff2b}) where $H/H_0$ can be
obtained from Eq. (\ref{ppa3}) with Eq. (\ref{ppa2}).

\begin{figure}[!ht]
\includegraphics[width=0.98\linewidth]{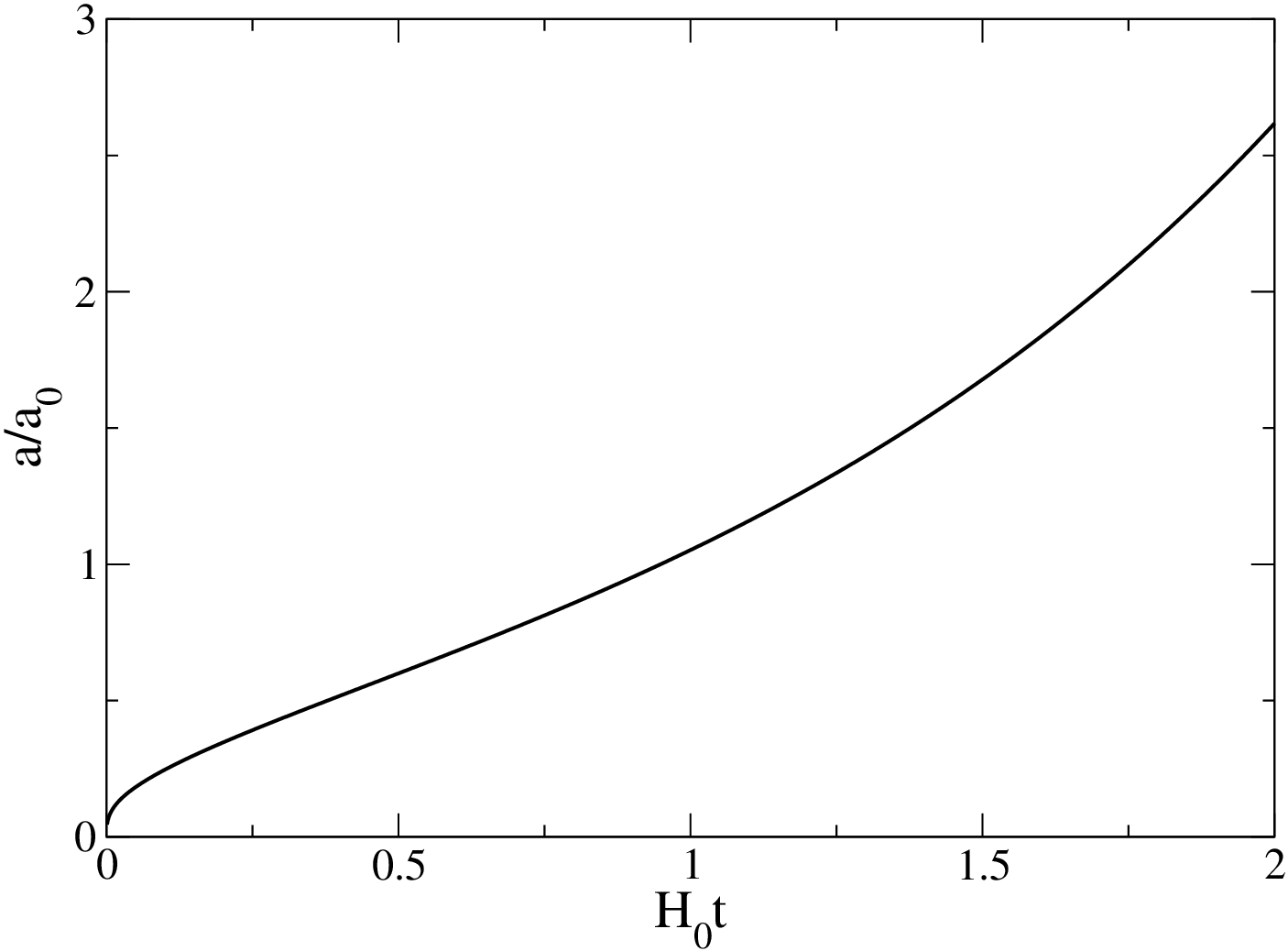}
\caption{Evolution of the scale factor as a function of time. \label{taPP}}
\end{figure}

\begin{figure}[!ht]
\includegraphics[width=0.98\linewidth]{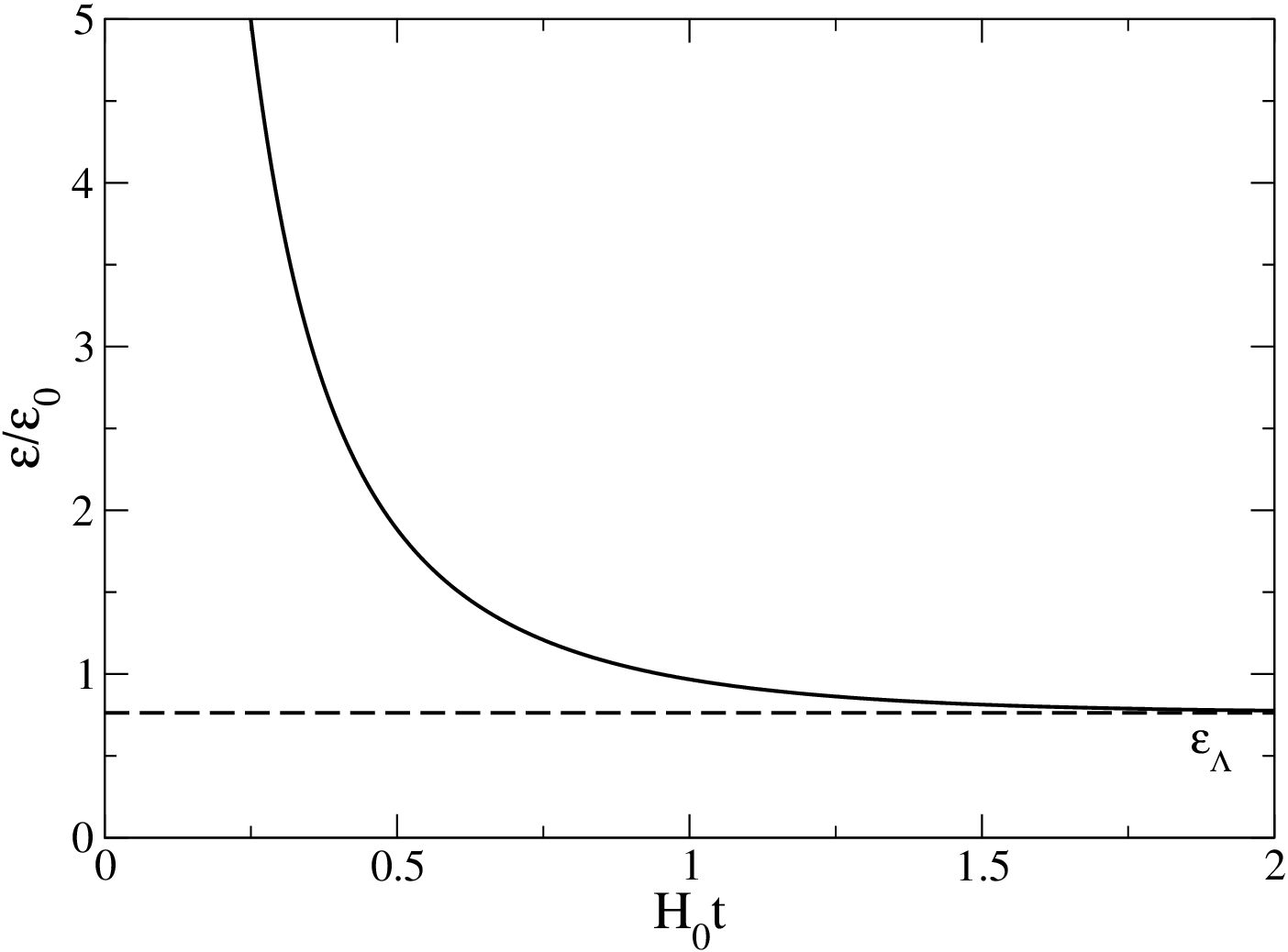}
\caption{Evolution of the energy density as a function of time. \label{tepsPP}}
\end{figure}

At $t=0$ the universe starts
from a singular state at
which the scale factor $a=0$
while the energy density $\epsilon=+\infty$.  The scale factor
increases with time. For $t\rightarrow
+\infty$,
we obtain
\begin{eqnarray}
\frac{a}{a_0}\sim \left (\frac{\Omega_{m,0}}{\Omega_{\Lambda,0}}+2\sqrt{\frac{\Omega_{s,0}}{\Omega_{\Lambda,0}}}\right )^{1/3}\frac{1}{2^{2/3}} e^{\sqrt{\Omega_{\Lambda,0}}H_0 t}.
\label{ppa4}
\end{eqnarray}
The energy density decreases
with time and tends to $\epsilon_{\Lambda}$ for $t\rightarrow +\infty$.
The expansion is decelerating during the stiff matter era and the dust matter
era while it is accelerating during the dark energy era. The
temporal evolutions of the scale factor and of the energy density are shown in
Figs. \ref{taPP} and \ref{tepsPP}.

\subsection{Stiff matter and dust matter}
\label{sec_ppb}

We consider a universe made of stiff matter and dust matter. In the absence of
dark energy ($\Omega_{\Lambda,0}=0$), using the identity
\begin{equation}
\int \frac{dx}{x\sqrt{\frac{a}{x^3}+\frac{b}{x^6}}}=\frac{2}{3a}\sqrt{b+ax^3},
\label{ppb1}
\end{equation}
we
obtain
\begin{eqnarray}
\frac{a}{a_0}=\left (\frac{9}{4}\Omega_{m,0}H_0^2 t^2+3\sqrt{\Omega_{s,0}}H_0 t\right )^{1/3},
\label{ppb2}
\end{eqnarray}
\begin{eqnarray}
\frac{\epsilon}{\epsilon_0}=\frac{4}{9H_0^2t^2}\left (\frac{1+\frac{2\sqrt{\Omega_{s,0}}}{3\Omega_{m,0}H_0 t}}{1+\frac{4\sqrt{\Omega_{s,0}}}{3\Omega_{m,0}H_0 t}}\right )^2.
\label{ppb3}
\end{eqnarray}

\subsection{Stiff matter and dark energy}
\label{sec_ppc}

We consider a universe made of stiff matter and dark energy. In the absence of matter ($\Omega_{m,0}=0$), using the identity
\begin{equation}
\int \frac{dx}{x\sqrt{\frac{b}{x^6}+c}}=\frac{1}{3\sqrt{c}}\ln\left\lbrack 2cx^3+2\sqrt{c}\sqrt{b+cx^6}\right\rbrack,
\label{ppc1}
\end{equation}
or setting $X=b/cx^6$ and using the identity
\begin{equation}
\int \frac{dX}{X\sqrt{X+1}}=\ln\left (\frac{\sqrt{1+X}-1}{\sqrt{1+X}+1}\right ),
\label{ppc2}
\end{equation}
we get
\begin{eqnarray}
\frac{a}{a_0}=\left (\frac{\Omega_{s,0}}{\Omega_{\Lambda,0}}\right )^{1/6}\sinh^{1/3}\left (3 \sqrt{\Omega_{\Lambda,0}}H_0 t\right ),
\label{ppc3}
\end{eqnarray}
\begin{eqnarray}
\frac{\epsilon}{\epsilon_0}=\frac{\Omega_{\Lambda,0}}{\tanh^2\left (3\sqrt{\Omega_{\Lambda,0}}H_0 t\right )}.
\label{ppc4}
\end{eqnarray}

\subsection{Stiff matter}
\label{sec_ppd}

We consider a universe made of stiff matter. In the absence of dust matter and
dark energy ($\Omega_{m,0}=\Omega_{\Lambda,0}=0$), we find that
\begin{eqnarray}
\frac{a}{a_0}=\left (3\sqrt{\Omega_{s,0}}H_0 t\right )^{1/3},\qquad \frac{\epsilon}{\epsilon_0}=\frac{1}{9H_0^2t^2}.
\label{ppd1}
\end{eqnarray}

\subsection{Dust matter and dark energy}
\label{sec_ppe}

We consider a universe made of dust matter and dark energy. 
In the absence of stiff matter ($\Omega_{s,0}=0$), using the identity
\begin{equation}
\int \frac{dx}{x\sqrt{\frac{a}{x^3}+c}}=\frac{1}{3\sqrt{c}}\ln\left\lbrack a+2cx^3+2\sqrt{c}\sqrt{ax^3+cx^6}\right\rbrack,
\label{ppe1}
\end{equation}
or setting $X=a/cx^3$ and using identity (\ref{ppc2}), we obtain
\begin{eqnarray}
\frac{a}{a_0}=\left (\frac{\Omega_{m,0}}{\Omega_{\Lambda,0}}\right )^{1/3}\sinh^{2/3}\left (\frac{3}{2}\sqrt{\Omega_{\Lambda,0}}H_0 t\right ),
\label{ppe2}
\end{eqnarray}
\begin{eqnarray}
\frac{\epsilon}{\epsilon_0}=\frac{\Omega_{\Lambda,0}}{\tanh^2\left (\frac{3}{2}\sqrt{\Omega_{\Lambda,0}}H_0 t\right )}.
\label{ppe3}
\end{eqnarray}
This corresponds to the $\Lambda$CDM
model.

\subsection{Dark energy}
\label{sec_ppf}

We consider a universe made of dark energy. In the absence of stiff matter and
dust matter  ($\Omega_{s,0}=\Omega_{m,0}=0$), we obtain
\begin{eqnarray}
a(t)=a(0)e^{\sqrt{\frac{\Lambda}{3}}t},\qquad
\epsilon=\epsilon_{\Lambda}.
\label{ppf1}
\end{eqnarray}
This is de Sitter's solution.

\subsection{Dust matter}
\label{sec_ppg}

We consider a universe made of dust matter. In the absence of stiff matter and
dark energy ($\Omega_{s,0}=\Omega_{\Lambda,0}=0$), we obtain 
\begin{eqnarray}
\frac{a}{a_0}=\left (\frac{9}{4}\Omega_{m,0}H_0^2 t^2\right )^{1/3},\qquad \frac{\epsilon}{\epsilon_0}=\frac{4}{9H_0^2t^2}.
\label{ppg1}
\end{eqnarray}
This is the Einstein-de Sitter (EdS) solution.

\section{The case $\Omega_{s,0}\le 0$ and $\Omega_{\Lambda,0}\ge 0$}
\label{sec_np}

\begin{figure}[!ht]
\includegraphics[width=0.98\linewidth]{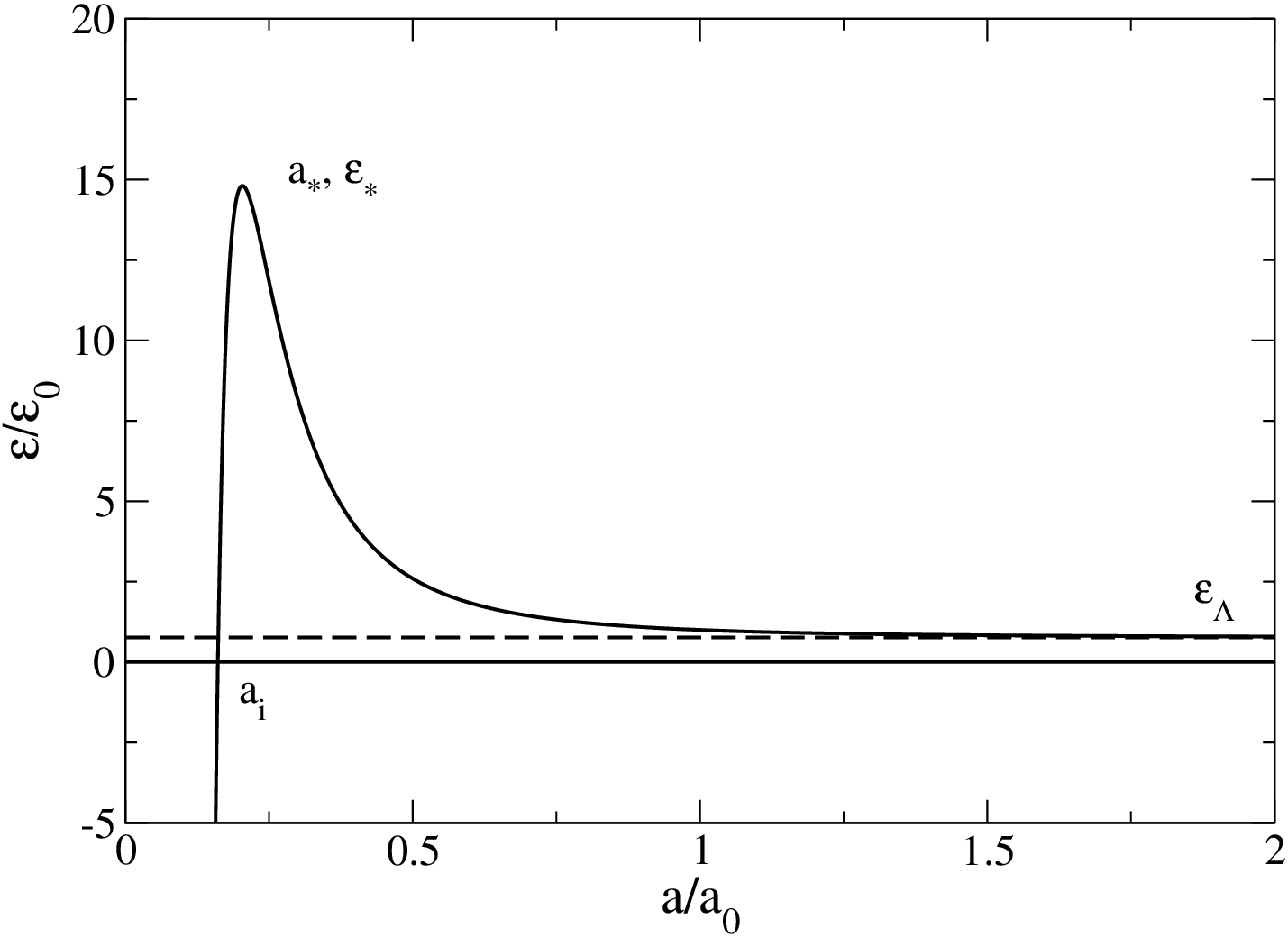}
\caption{Energy density as a function of the scale factor. We have taken
$\Omega_{m,0}=0.237$, $\Omega_{\Lambda,0}=0.763$, and
$\Omega_{s,0}=-10^{-3}$.\label{aepsNP}}
\end{figure}

\begin{figure}[!ht]
\includegraphics[width=0.98\linewidth]{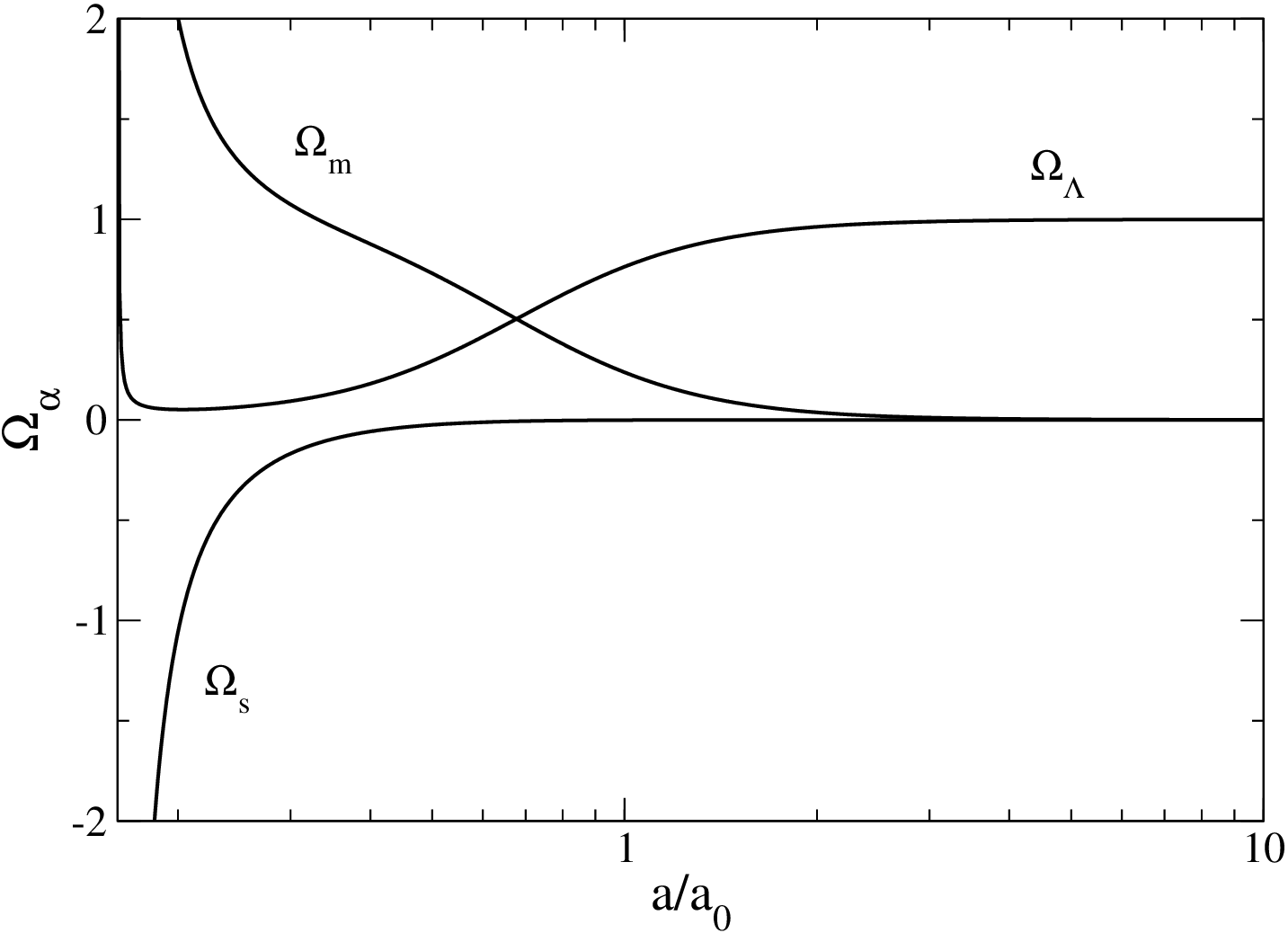}
\caption{Evolution of the proportion
$\Omega_{\alpha}=\epsilon_{\alpha}/\epsilon$ of the different components of
the universe with
the scale factor. \label{proportionsNP}}
\end{figure}

We consider the case of a negative stiff energy density
($\Omega_{s,0}\le 0$) and a positive cosmological constant
($\Omega_{\Lambda,0}\ge
0$). The total energy density is
\begin{equation}
\frac{\epsilon}{\epsilon_0}=-\frac{|\Omega_{s,0}|}{(a/a_0)^6}+\frac{\Omega_{m,0}}{(a/a_0)^3}+\Omega_{\Lambda,0}.
\label{np1}
\end{equation}
The energy density is positive for $a\ge a_i$ with 
\begin{equation}
\frac{a_i}{a_0}=\left
(\frac{-\Omega_{m,0}+\sqrt{\Delta}}{2\Omega_{\Lambda,0}}\right
)^{1/3},
\label{np2}
\end{equation}
where we have defined
\begin{eqnarray}
\Delta=\Omega^2_{m,0}+4\Omega_{\Lambda,0}|\Omega_{s,0}|.
\label{np2b}
\end{eqnarray}
The energy density starts from $\epsilon=0$ at $a=a_i$, increases, reaches a
maximum at
\begin{equation}
\frac{a_*}{a_0}=\left (\frac{2|\Omega_{s,0}|}{\Omega_{m,0}}\right )^{1/3},\quad
\frac{\epsilon_*}{\epsilon_0}=\frac{\Delta}{4|\Omega_{s,0}|},
\label{np3}
\end{equation}
decreases, and tends to $\epsilon_{\Lambda}$ for $a\rightarrow +\infty$. The
relation between the energy density and the scale factor is shown
in Fig. \ref{aepsNP}. The proportions of stiff matter, dust matter and dark
energy as a function  of the scale factor are shown in Fig.
\ref{proportionsNP}.

\subsection{Anti-stiff matter, dust matter, and dark energy}
\label{sec_npa}

We consider a universe made of anti-stiff matter, dust matter, and dark energy.
Using the identity (\ref{ppa1}), Eq. (\ref{stiff3c}) with $a_i$ given by Eq.
(\ref{np2}) can be solved analytically to give 
\begin{eqnarray}
\frac{a}{a_0}=\left\lbrack \frac{\sqrt{\Delta}}{2\Omega_{\Lambda,0}}\cosh\left
(3\sqrt{\Omega_{\Lambda,0}} H_0 t\right
)-\frac{\Omega_{m,0}}{2\Omega_{\Lambda,0}}\right\rbrack^{1/3}.
\label{npa1}
\end{eqnarray}
From Eq. (\ref{npa1}), we can compute $H=\dot a/a$ leading to
\begin{eqnarray}
\left (\frac{a}{a_0}\right
)^3\frac{H}{H_0}=\frac{\sqrt{\Delta}}{2\sqrt{\Omega_{\Lambda,0}}}\sinh\left
(3\sqrt{\Omega_{\Lambda,0}} H_0 t\right
).
\label{npa2b}
\end{eqnarray}
The energy density $\epsilon/\epsilon_0$ is given by Eq. (\ref{stiff2b}) where
$H/H_0$ can be obtained from Eq.  (\ref{npa2b}) with Eq. (\ref{npa1}).

\begin{figure}[!ht]
\includegraphics[width=0.98\linewidth]{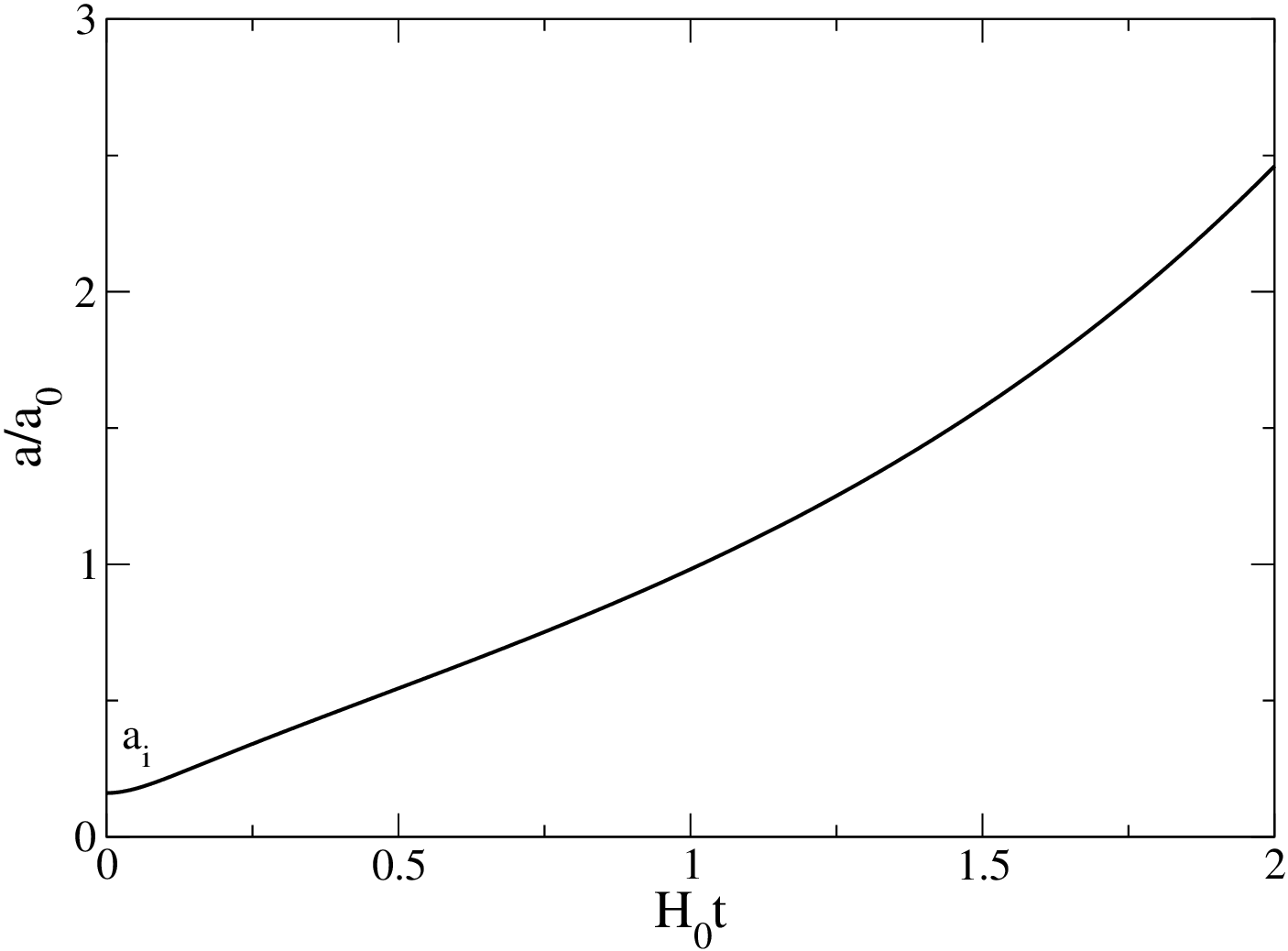}
\caption{Evolution of the scale factor as a function of time. \label{taNP}}
\end{figure}

\begin{figure}[!ht]
\includegraphics[width=0.98\linewidth]{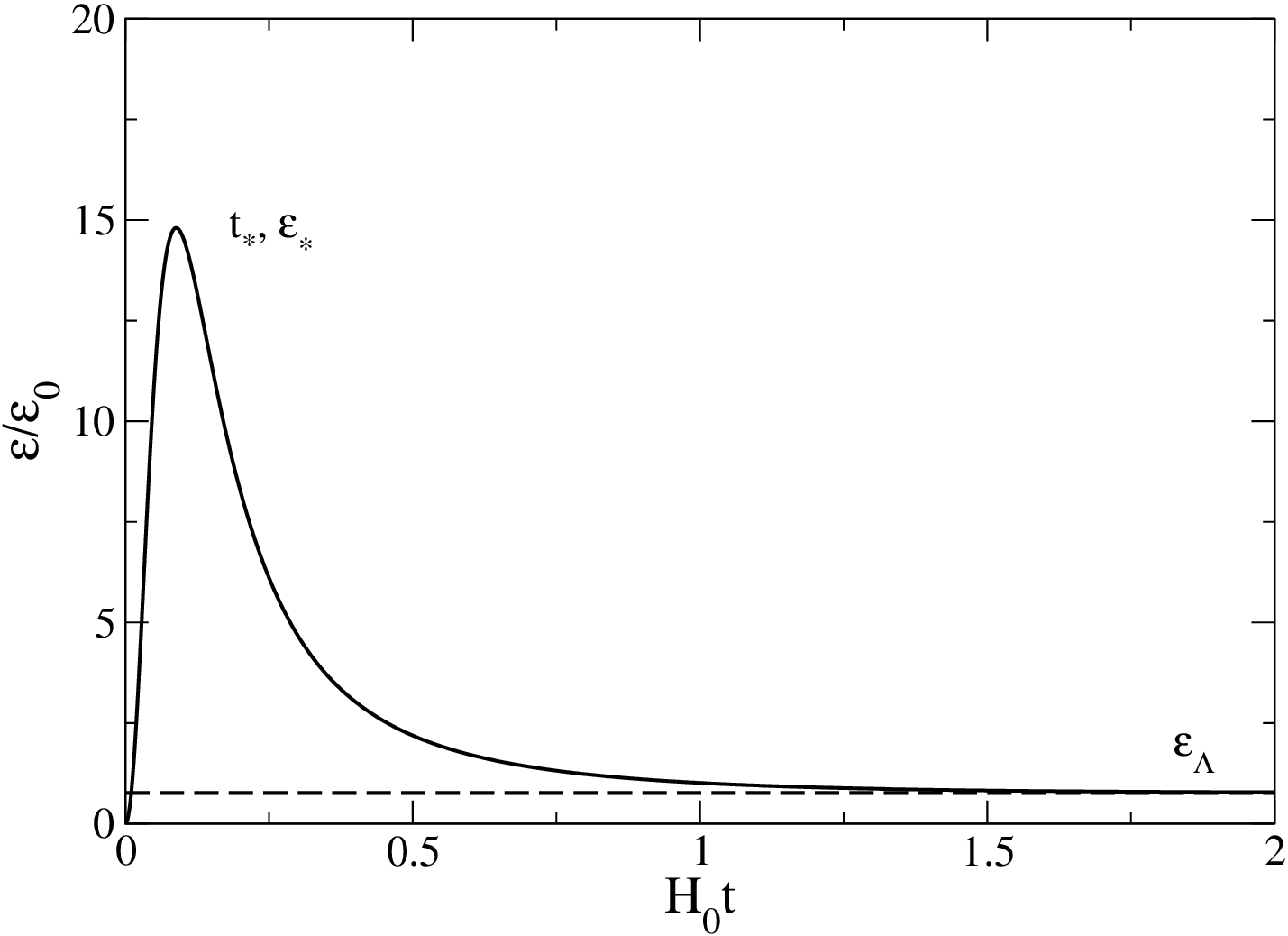}
\caption{Evolution of the energy density as a function of time.\label{tepsNP}}
\end{figure}

At $t=0$ the universe starts from a non-singular state at which the scale factor
$a=a_i$ and the energy density $\epsilon=0$. The scale factor increases with
time. For $t\rightarrow +\infty$, we obtain
\begin{eqnarray}
\frac{a}{a_0}\sim \left \lbrack \left (\frac{\Omega_{m,0}}{\Omega_{\Lambda,0}}\right )^2+4\frac{|\Omega_{s,0}|}{\Omega_{\Lambda,0}}\right \rbrack^{1/6}\frac{1}{2^{2/3}} e^{\sqrt{\Omega_{\Lambda,0}}H_0 t}.
\label{npa3}
\end{eqnarray}
The energy density increases, reaches its maximum value
$\epsilon_*$ at $t=t_*$ where
\begin{eqnarray}
t_*=\frac{1}{3\sqrt{\Omega_{\Lambda,0}}H_0}\ln\left
(\frac{\sqrt{\Delta}+\sqrt{4\Omega_{\Lambda,0}|\Omega_{s,0}|}}{\Omega_{m,0}}
\right ),
\label{npa4}
\end{eqnarray}
decreases and tends to $\epsilon_{\Lambda}$ for
$t\rightarrow
+\infty$. The universe is accelerating during the anti-stiff matter era,
decelerating during the dust matter era, and accelerating during the dark
matter era. The temporal evolutions of the scale factor and of the energy
density
are shown in Figs. \ref{taNP} and \ref{tepsNP}.

\subsection{Anti-stiff matter and dust matter}
\label{sec_npb}

We consider a universe made of anti-stiff matter and dust matter.  In the
absence of dark energy ($\Omega_{\Lambda,0}=0$), using the identity
(\ref{ppb1}), we obtain
\begin{eqnarray}
\frac{a}{a_0}=\left (\frac{9}{4}\Omega_{m,0}H_0^2 t^2+\frac{|\Omega_{s,0}|}{\Omega_{m,0}}\right )^{1/3},
\label{npb1}
\end{eqnarray}
\begin{eqnarray}
\frac{\epsilon}{\epsilon_0}=\frac{4}{9H_0^2t^2}\frac{1}{\left ( 1+\frac{4|\Omega_{s,0}|}{9\Omega^2_{m,0} H_0^2 t^2}\right )^2}.
\label{npb2}
\end{eqnarray}
At $t=0$ the universe starts from a non-singular state  at which the scale factor $a=a_i$ with
\begin{eqnarray}
\frac{a_i}{a_0}=\left (\frac{|\Omega_{s,0}|}{\Omega_{m,0}}\right )^{1/3},
\label{npb3}
\end{eqnarray}
and the energy density $\epsilon=0$. The scale factor increases
with time. The energy density increases,
reaches its maximum value
\begin{equation}
\frac{a_*}{a_0}=\left (\frac{2|\Omega_{s,0}|}{\Omega_{m,0}}\right )^{1/3},\quad \frac{\epsilon_*}{\epsilon_0}=\frac{\Omega_{m,0}^2}{4|\Omega_{s,0}|},
\label{npb4}
\end{equation}
at
\begin{eqnarray}
t_*=\frac{2\sqrt{|\Omega_{s,0}|}}{3\Omega_{m,0}H_0},
\label{npb5}
\end{eqnarray}
and decreases.

\subsection{Anti-stiff matter and dark energy}
\label{sec_npc}

We consider a universe made of anti-stiff matter and dark energy. In the absence
of matter ($\Omega_{m,0}=0$), 
using the identities (\ref{ppc1}) and (\ref{ppc2}), we obtain
\begin{eqnarray}
\frac{a}{a_0}=\left (\frac{|\Omega_{s,0}|}{\Omega_{\Lambda,0}}\right )^{1/6}\cosh^{1/3}\left (3 \sqrt{\Omega_{\Lambda,0}}H_0 t\right ),
\label{npc1}
\end{eqnarray}
\begin{eqnarray}
\frac{\epsilon}{\epsilon_0}=\Omega_{\Lambda,0}{\tanh^2\left (3\sqrt{\Omega_{\Lambda,0}}H_0 t\right )}.
\label{npc2}
\end{eqnarray}
At $t=0$ the universe starts from a non-singular state  at which the scale
factor $a=a_i$ with
\begin{eqnarray}
\frac{a_i}{a_0}=\left (\frac{|\Omega_{s,0}|}{\Omega_{\Lambda,0}}\right )^{1/6},
\label{npc3}
\end{eqnarray}
and the energy density $\epsilon=0$. The scale factor increases 
with time. The energy density increases with time and tends to 
$\epsilon_{\Lambda}$ for $t\rightarrow
+\infty$.

\section{The case $\Omega_{s,0}\ge 0$ and $\Omega_{\Lambda,0}\le 0$}
\label{sec_pn}

We consider the case of a positive stiff energy density ($\Omega_{s,0}\ge 0$)
and a negative cosmological constant ($\Omega_{\Lambda,0}\le 0$). The total
energy density is
\begin{equation}
\frac{\epsilon}{\epsilon_0}=\frac{\Omega_{s,0}}{(a/a_0)^6}+\frac{\Omega_{m,0}}{(a/a_0)^3}
-|\Omega_{\Lambda,0}|.
\label{pn1}
\end{equation}
The energy density is positive for $a\le a_f$ with
\begin{equation}
\frac{a_f}{a_0}=\left
(\frac{\Omega_{m,0}+\sqrt{\Delta}}{2|\Omega_{\Lambda,0}|}\right
)^{1/3},
\label{pn2}
\end{equation}
where we have defined
\begin{eqnarray}
\Delta=\Omega^2_{m,0}+4|\Omega_{\Lambda,0}|\Omega_{s,0}.
\label{pn2b}
\end{eqnarray}

\begin{figure}[!ht]
\includegraphics[width=0.98\linewidth]{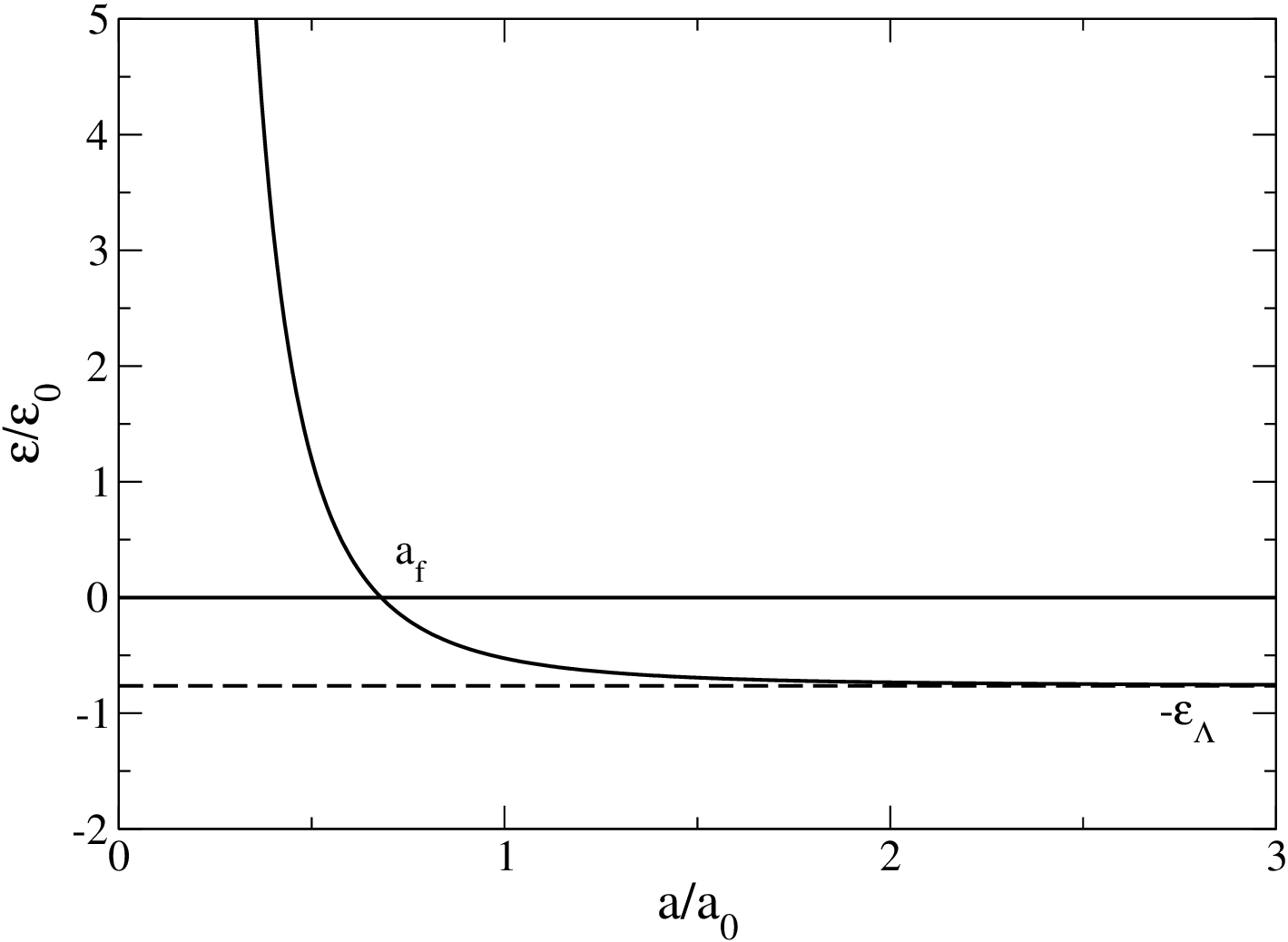}
\caption{Energy density as a function of the scale factor. We have taken
$\Omega_{m,0}=0.237$, $\Omega_{\Lambda,0}=-0.763$, and
$\Omega_{s,0}=10^{-3}$.
\label{aepsPN}}
\end{figure}

\begin{figure}[!ht]
\includegraphics[width=0.98\linewidth]{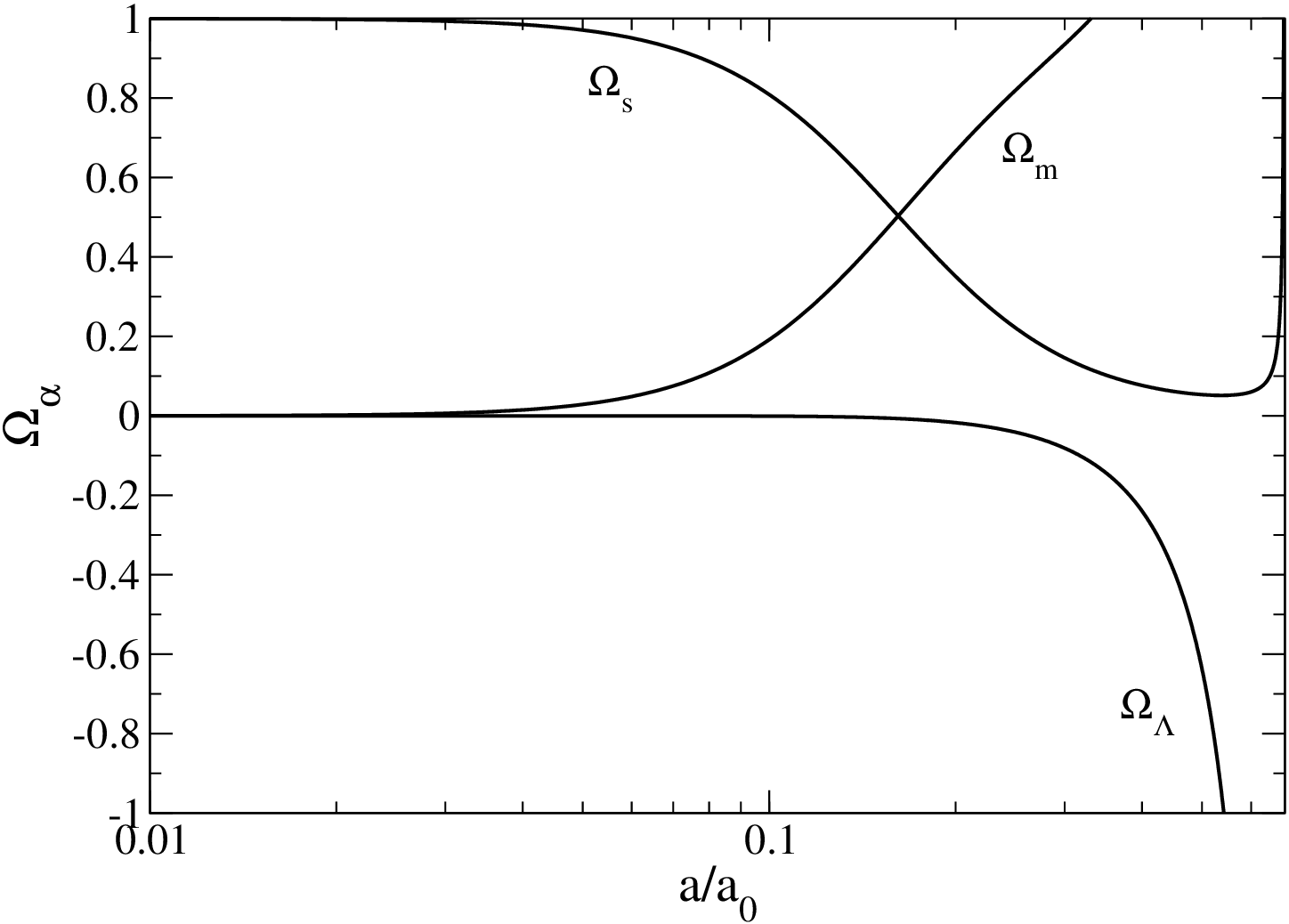}
\caption{Evolution of the proportion
$\Omega_{\alpha}=\epsilon_{\alpha}/\epsilon$ of the different components of
the universe with
the scale factor. \label{proportionsPN}}
\end{figure}

The energy density starts from $\epsilon=+\infty$ at $a=0$, decreases, and
reaches $\epsilon=0$ at $a=a_f$. The
relation between the energy density and the scale factor is shown
in Fig. \ref{aepsPN}. The proportions of stiff matter, dust matter and dark
energy as a function of the scale factor are shown in Fig.
\ref{proportionsPN}.

\subsection{Stiff matter, dust matter, and anti-dark energy}
\label{sec_pna}

We consider a universe made of stiff matter, dust matter, and anti-dark energy.
From Eqs. (\ref{ppa2}) and (\ref{ppa3}) with $\Omega_{\Lambda,0}<0$ we get 
\begin{eqnarray}
\frac{a}{a_0}=\Biggl \lbrack
\frac{\Omega_{m,0}}{|\Omega_{\Lambda,0}|}\sin^2\left
(\frac{3}{2}\sqrt{|\Omega_{\Lambda,0}|}H_0 t\right )\nonumber\\
+\sqrt{\frac{\Omega_{s,0}}{|\Omega_{\Lambda,0}|}}\sin \left
(3\sqrt{|\Omega_{\Lambda,0}|}H_0 t\right )\Biggr\rbrack^{1/3},
\label{pna1}
\end{eqnarray}
and 
\begin{eqnarray}
\left (\frac{a}{a_0}\right
)^3\frac{H}{H_0}=\frac{\Omega_{m,0}}{2\sqrt{|\Omega_{\Lambda,0}|}}\sin\left
(3\sqrt{|\Omega_{\Lambda,0}|}H_0 t\right
)\nonumber\\
+\sqrt{\Omega_{s,0}}\cos\left (3 \sqrt{|\Omega_{\Lambda,0}|}H_0
t\right ).
\label{pna1b}
\end{eqnarray}
The energy density is given by Eq. (\ref{stiff2b}) where $H/H_0$ can be
obtained from Eq. (\ref{pna1b}) with Eq. (\ref{pna1}). 

\begin{figure}[!ht]
\includegraphics[width=0.98\linewidth]{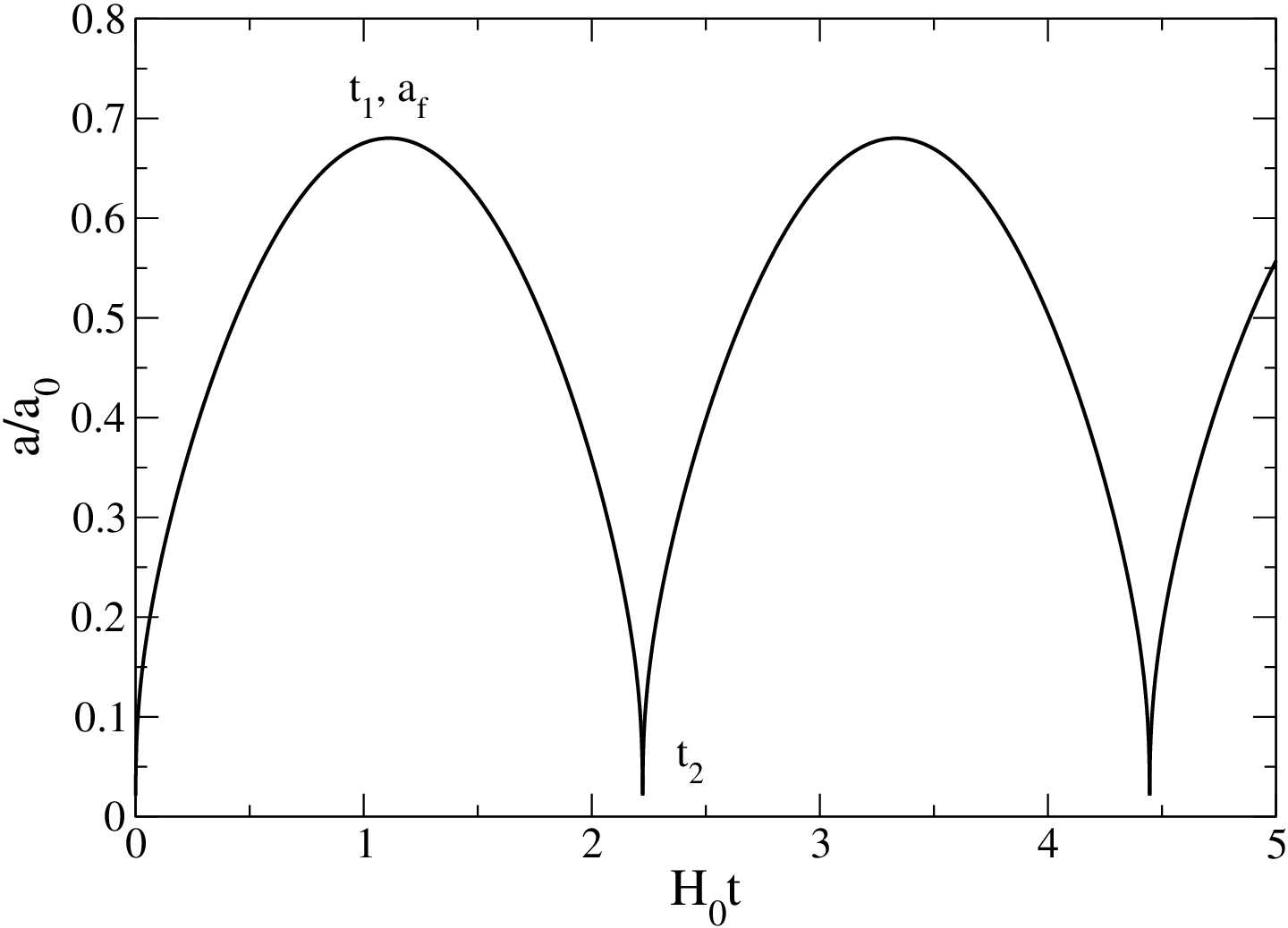}
\caption{Evolution of the scale factor as a function of time.
\label{taPN}}
\end{figure}

\begin{figure}[!ht]
\includegraphics[width=0.98\linewidth]{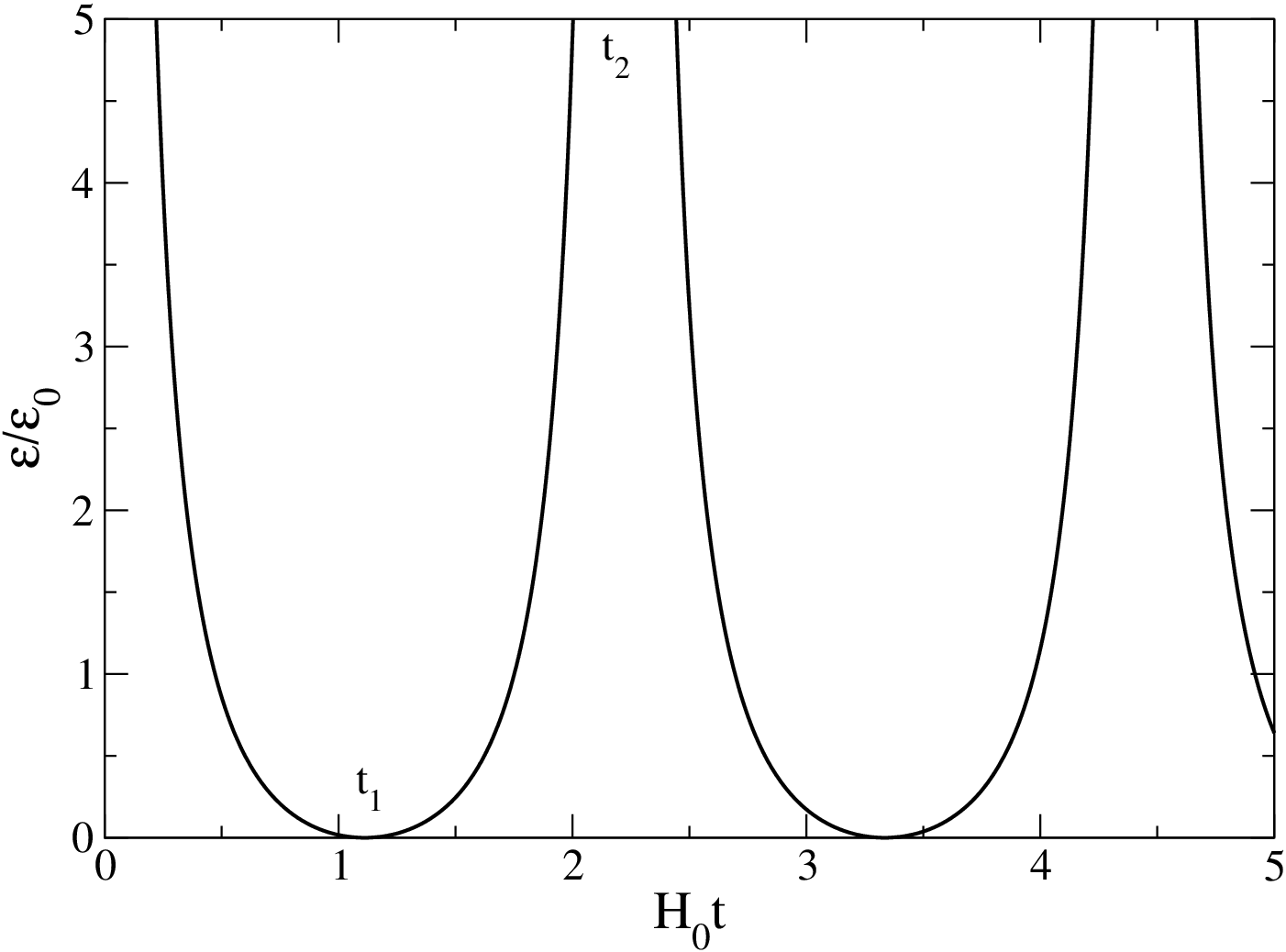}
\caption{Evolution of the energy density as a function of time.\label{tepsPN}}
\end{figure}

At $t=0$ the universe starts
from a singularity at which the scale factor $a=0$ and the energy density
$\epsilon=+\infty$.  Between $t=0$ and $t=t_1$ where
\begin{eqnarray}
t_1=\frac{\pi-\tan^{-1}(2\sqrt{\Omega_{s,0}|\Omega_{\Lambda,0}}|/\Omega_{m,0})}{3\sqrt{|\Omega_{\Lambda,0}|}H_0},
\label{pna2}
\end{eqnarray}
the energy density decreases from $\epsilon=+\infty$ to $\epsilon=0$ and the
scale factor increases from $a=0$ to $a=a_f$. Between $t=t_1$ and $t=t_2=2t_1$
the energy density increases from $\epsilon=0$ to $\epsilon=+\infty$ and the
scale factor decreases from $a=a_f$ to $a=0$. This process continues
periodically with a period $t_2$. The temporal evolutions of the scale factor
and of the energy density are shown in Figs. \ref{taPN} and \ref{tepsPN}.

\subsection{Stiff matter and anti-dark energy}
\label{sec_pnb}

We consider a universe made of stiff matter and anti-dark energy. In the absence of matter ($\Omega_{m,0}=0$), we get
\begin{eqnarray}
\frac{a}{a_0}=\left (\frac{\Omega_{s,0}}{|\Omega_{\Lambda,0}|}\right )^{1/6}\sin^{1/3}\left (3 \sqrt{|\Omega_{\Lambda,0}|}H_0 t\right ),
\label{pnb1}
\end{eqnarray}
\begin{eqnarray}
\frac{\epsilon}{\epsilon_0}=\frac{|\Omega_{\Lambda,0}|}{\tan^2\left (3\sqrt{|\Omega_{\Lambda,0}|}H_0 t\right )}.
\label{pnb2}
\end{eqnarray}
At $t=0$ the universe starts from a singularity at which the scale factor $a=0$
and the energy density $\epsilon=+\infty$. Between $t=0$
and $t=t_1$ where
\begin{eqnarray}
t_1=\frac{\pi}{6\sqrt{|\Omega_{\Lambda,0}|}H_0},
\label{pnb3}
\end{eqnarray}
the energy density decreases from $\epsilon=+\infty$ to $\epsilon=0$ and the scale factor increases from $a=0$ to $a=a_f$ where
\begin{equation}
\frac{a_f}{a_0}=\left (\frac{\Omega_{s,0}}{|\Omega_{\Lambda,0}|}\right )^{1/6}.
\label{pnb4}
\end{equation}
Between $t=t_1$ and $t=t_2=2t_1$
the energy density increases from $\epsilon=0$ to $\epsilon=+\infty$ and the
scale factor decreases from $a=a_f$ to $a=0$. This process continues
periodically with a period $t_2$.

\subsection{Dust matter and anti-dark energy}
\label{sec_pnc}

We consider a universe made of dust matter and anti-dark energy.  In the absence
of stiff matter ($\Omega_{s,0}=0$), we obtain
\begin{eqnarray}
\frac{a}{a_0}=\left (\frac{\Omega_{m,0}}{|\Omega_{\Lambda,0}|}\right )^{1/3}\sin^{2/3}\left (\frac{3}{2}\sqrt{|\Omega_{\Lambda,0}|}H_0 t\right ),
\label{pnc1}
\end{eqnarray}
\begin{eqnarray}
\frac{\epsilon}{\epsilon_0}=\frac{|\Omega_{\Lambda,0}|}{\tan^2\left (\frac{3}{2}\sqrt{|\Omega_{\Lambda,0}|}H_0 t\right )}.
\label{pnc2}
\end{eqnarray}
At $t=0$ the universe starts from a singularity at which the scale factor $a=0$
and the energy density $\epsilon=+\infty$. Between $t=0$
and $t=t_1$ where
\begin{eqnarray}
t_1=\frac{\pi}{3\sqrt{|\Omega_{\Lambda,0}|}H_0},
\label{pnc3}
\end{eqnarray}
the energy density decreases from $\epsilon=+\infty$ to $\epsilon=0$ and the scale factor increases from $a=0$ to $a=a_f$ where
\begin{equation}
\frac{a_f}{a_0}=\left (\frac{\Omega_{m,0}}{|\Omega_{\Lambda,0}|}\right )^{1/3}.
\label{pnc4}
\end{equation}
Between $t=t_1$ and $t=t_2=2t_1$ 
the energy density increases from $\epsilon=0$ to $\epsilon=+\infty$ and
the scale factor decreases from $a=a_f$ to $a=0$. This process continues
periodically with a period $t_2$.
This solution  corresponds to
the anti-$\Lambda$CDM model. 

\section{The case $\Omega_{s,0}\le 0$ and $\Omega_{\Lambda,0}\le 0$}
\label{sec_nn}

We consider the case of a negative stiff energy density ($\Omega_{s,0}\le 0$)
and a negative cosmological constant ($\Omega_{\Lambda,0}\le 0$). The total
energy density is
\begin{equation}
\frac{\epsilon}{\epsilon_0}=-\frac{|\Omega_{s,0}|}{(a/a_0)^6}+\frac{\Omega_{m,0}}{(a/a_0)^3}
-|\Omega_{\Lambda,0}|.
\label{nn1}
\end{equation}
If $|\Omega_{s,0}|>\Omega^2_{m,0}/(4|\Omega_{\Lambda,0}|)$ the energy density is always negative so this situation is not possible. Therefore, we assume $|\Omega_{s,0}|\le\Omega^2_{m,0}/(4|\Omega_{\Lambda,0}|)$.
In that case, the energy density is positive for $a_i\le a\le a_f$ with
\begin{equation}
\frac{a_i}{a_0}=\left
(\frac{\Omega_{m,0}-\sqrt{\Delta}}{2|\Omega_{\Lambda,0}|}\right
)^{1/3},
\label{nn2}
\end{equation}
and
\begin{equation}
\frac{a_f}{a_0}=\left
(\frac{\Omega_{m,0}+\sqrt{\Delta}}{2|\Omega_{\Lambda,0}|}\right
)^{1/3},
\label{nn3}
\end{equation}
where we have defined
\begin{eqnarray}
\Delta=\Omega^2_{m,0}-4|\Omega_{\Lambda,0}||\Omega_{s,0}|.
\label{nna2}
\end{eqnarray}

\begin{figure}[!ht]
\includegraphics[width=0.98\linewidth]{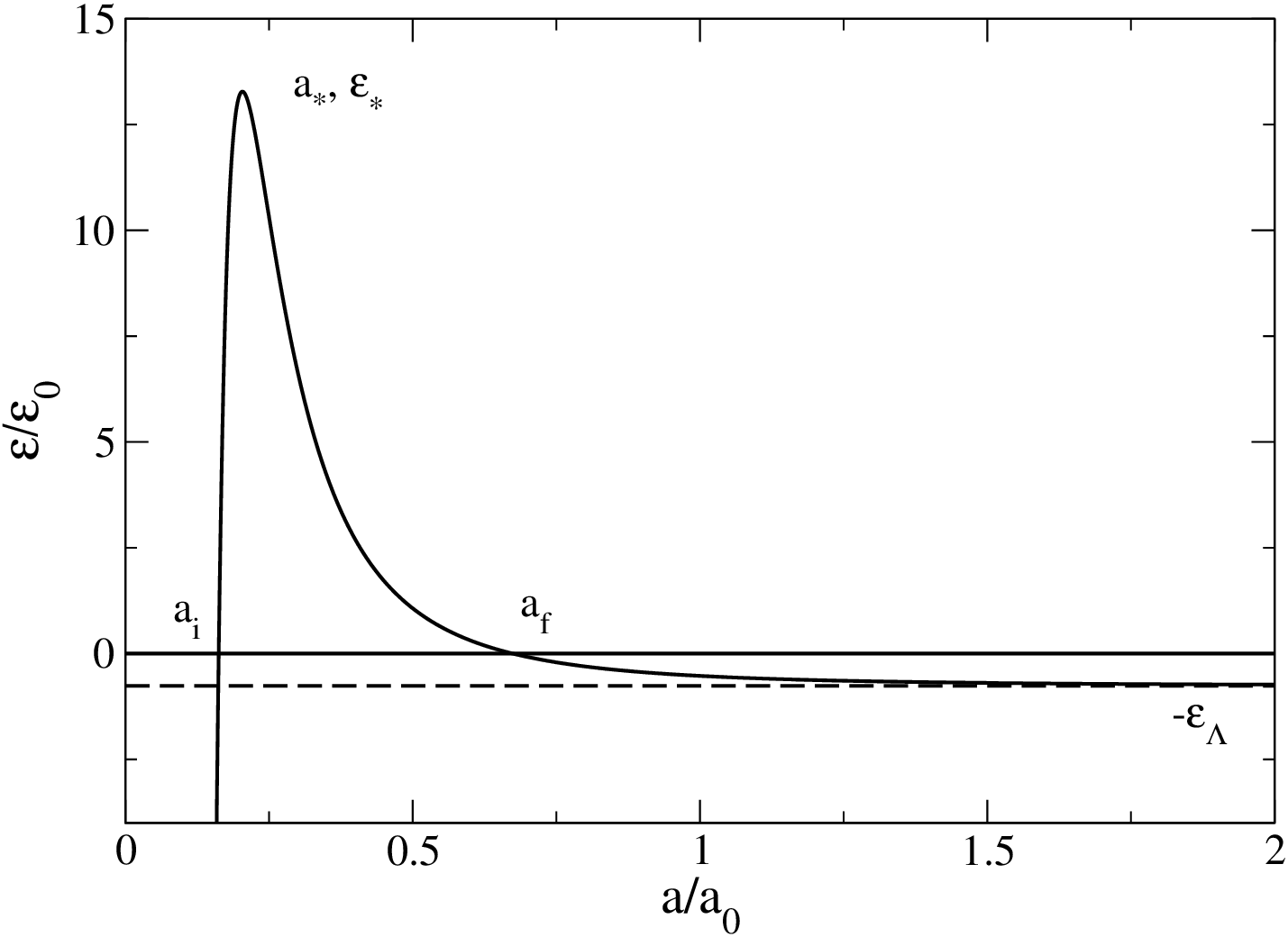}
\caption{Energy density as a function of the scale factor. We have taken
$\Omega_{m,0}=0.237$, $\Omega_{\Lambda,0}=-0.763$, and
$\Omega_{s,0}=-10^{-3}$.\label{aepsNN}}
\end{figure}

\begin{figure}[!ht]
\includegraphics[width=0.98\linewidth]{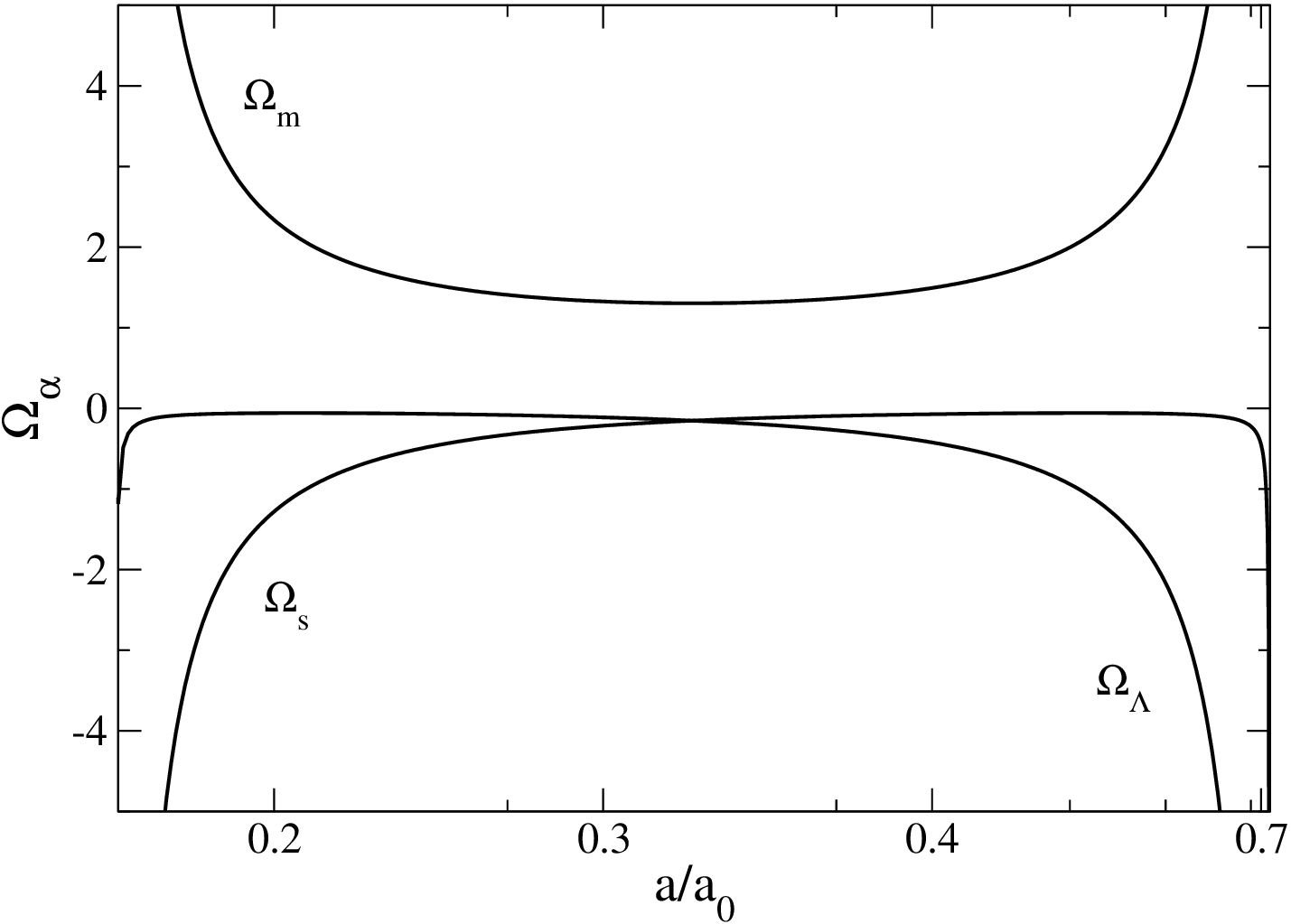}
\caption{Evolution of the proportion
$\Omega_{\alpha}=\epsilon_{\alpha}/\epsilon$ of the different components of
the universe with
the scale factor. \label{proportionsNN}}
\end{figure}

The energy density starts from $\epsilon=0$ at $a=a_i$, increases, reaches a maximum at
\begin{equation}
\frac{a_*}{a_0}=\left (\frac{2|\Omega_{s,0}|}{\Omega_{m,0}}\right )^{1/3},\quad
\frac{\epsilon_*}{\epsilon_0}=\frac{\Delta}{4|\Omega_{s,0}|},
\label{nn4}
\end{equation}
decreases, and reaches $\epsilon=0$ at $a=a_f$. The
relation between the energy density and the scale factor is shown
in Fig. \ref{aepsNN}. The proportions of stiff matter, dust matter and dark
energy as a function of the scale factor are shown in Fig.
\ref{proportionsNN}.

\subsection{Anti-stiff matter, dust matter, and anti-dark energy}
\label{sec_nna}

We consider a universe made of anti-stiff matter, dust matter, and anti-dark
energy. From  Eqs. (\ref{npa1}) and (\ref{npa2b}) with $\Omega_{\Lambda,0}<0$,
we get
\begin{eqnarray}
\frac{a}{a_0}=\left\lbrack
\frac{\Omega_{m,0}}{2|\Omega_{\Lambda,0}|}-\frac{\sqrt{\Delta}}{2|\Omega_{
\Lambda,0}|}\cos \left (3\sqrt{|\Omega_{\Lambda,0}|}H_0 t\right
)\right\rbrack^{1/3},
\label{nna1}
\end{eqnarray}
and
\begin{eqnarray}
\left (\frac{a}{a_0}\right
)^3\frac{H}{H_0}=\frac{\sqrt{\Delta}}{2\sqrt{|\Omega_{\Lambda,0}|}}\sin\left
(3\sqrt{|\Omega_{\Lambda,0}|}H_0 t\right
).
\label{nna2b}
\end{eqnarray}
The energy density is given by Eq. (\ref{stiff2b}) where $H/H_0$ can be
obtained from Eq.  (\ref{nna2b}) with Eq. (\ref{nna1}).

\begin{figure}[!ht]
\includegraphics[width=0.98\linewidth]{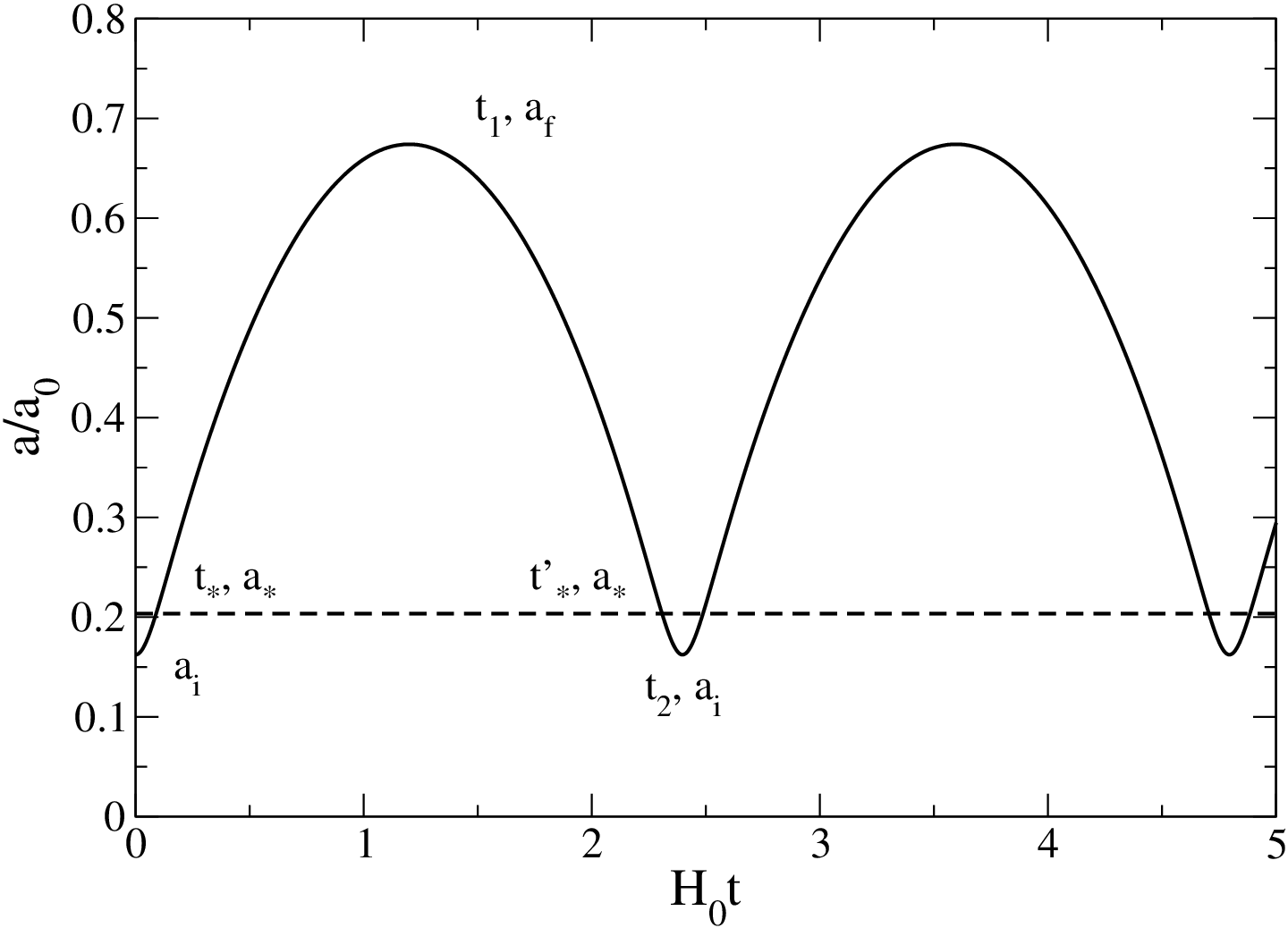}
\caption{Evolution of the scale factor as a function of time.\label{taNN}}
\end{figure}

\begin{figure}[!ht]
\includegraphics[width=0.98\linewidth]{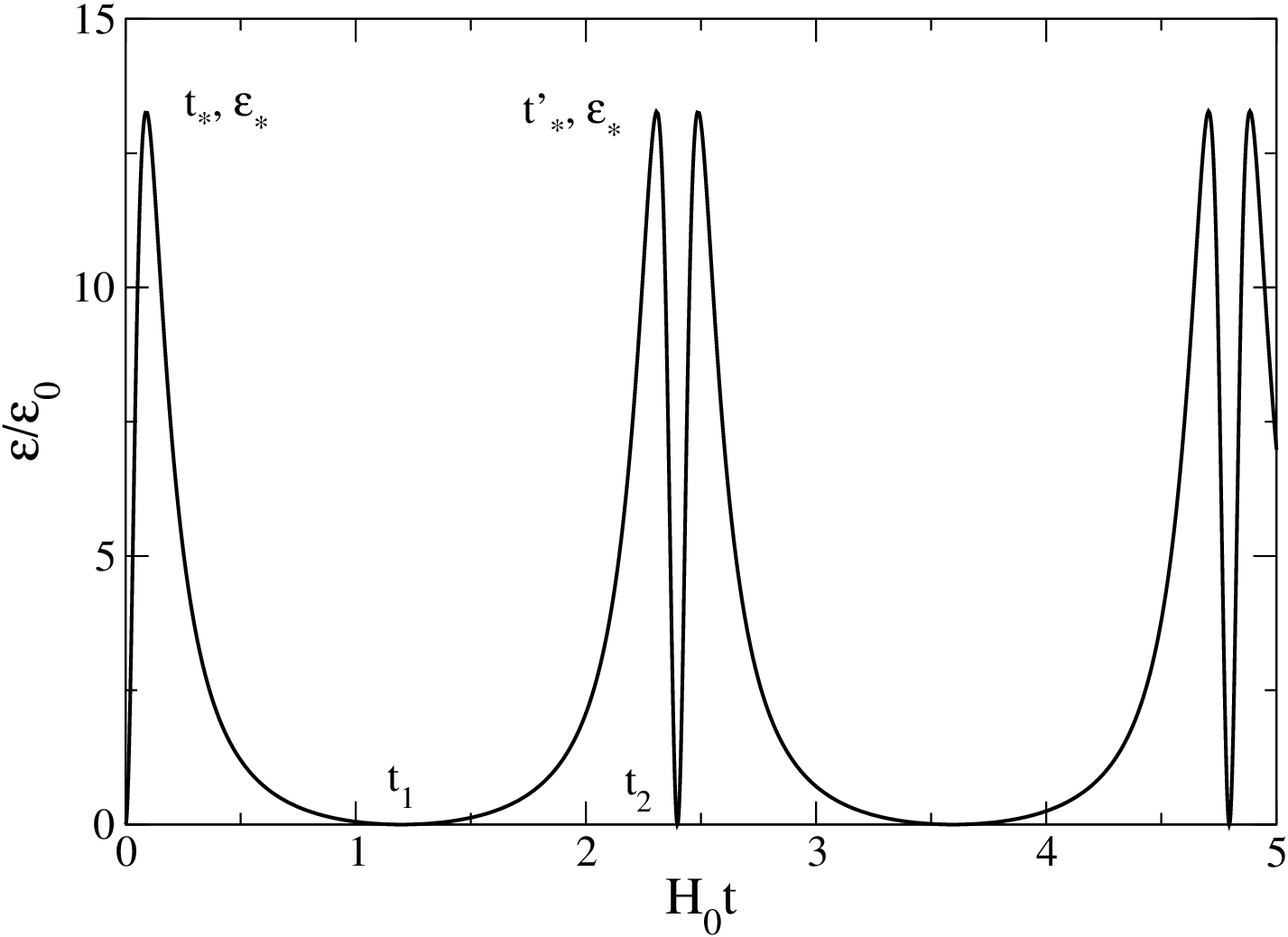}
\caption{Evolution of the energy density as a function of time.\label{tepsNN}}
\end{figure}

At $t=0$ the universe starts from a non-singular state at which the scale factor $a=a_i$ and the energy density $\epsilon=0$. Between $t=0$ and $t=t_*$ where
\begin{eqnarray}
t_*=\frac{\cos^{-1}(\sqrt{\Delta}/\Omega_{m,0})}{3\sqrt{|\Omega_{\Lambda,0}|}H_0},
\label{nna3}
\end{eqnarray}
the energy density increases from $\epsilon=0$ to its
maximum value $\epsilon=\epsilon_*$ and the scale factor increases from $a=a_i$
to $a=a_*$. Between $t=t_*$ and $t=t_1$ where
\begin{eqnarray}
t_1=\frac{\pi}{3\sqrt{|\Omega_{\Lambda,0}|}H_0},
\label{nna4}
\end{eqnarray}
the energy density decreases from $\epsilon=\epsilon_*$ to $\epsilon=0$ and the scale factor increases from $a=a_*$ to $a=a_f$. Between $t=t_1$ and $t'_*$ where
\begin{eqnarray}
t'_*=\frac{2\pi-\cos^{-1}(\sqrt{\Delta}/\Omega_{m,0})}{3\sqrt{|\Omega_{\Lambda,0}|}H_0},
\label{nna5}
\end{eqnarray}
the energy density increases from $\epsilon=0$ to $\epsilon=\epsilon_*$ and the
scale factor decreases from $a=a_f$ to $a=a_*$.  Between $t'_*$ and $t_2=2 t_1$
the energy density decreases from $\epsilon=\epsilon_*$ to $\epsilon=0$ and the
scale factor decreases from $a=a_*$ to $a=a_i$. This process continues
periodically with a period $t_2$.
The temporal evolutions of the scale factor
and of the energy density are shown in Figs. \ref{taNN} and \ref{tepsNN}.

\section{Some analytical solutions including the radiation era}
\label{sec_rad}

We now come back to the general equation (\ref{stiff3}) including the
contribution of radiation and provide some particular analytical solutions.

\subsection{Stiff matter and radiation}
\label{sec_rada}

We consider  a  universe made of stiff matter and radiation. The total energy is
\begin{equation}
\frac{\epsilon}{\epsilon_0}=\frac{\Omega_{s,0}}{(a/a_0)^6}+\frac{\Omega_{rad,0}
}{(a/a_0)^4}.
\label{rada0}
\end{equation}
The energy density starts from $\epsilon=+\infty$ at $a=a_i=0$ and decreases
when $a$ increases. In the absence of dust matter and dark energy
($\Omega_{m,0}=\Omega_{\Lambda,0}=0$) the integral in Eq. (\ref{stiff3})
can be performed analytically giving
\begin{eqnarray}
2\sqrt{\Omega_{rad,0}}\frac{a}{a_0}\sqrt{\Omega_{s,0}+\Omega_{rad,0}\left
(\frac{a}{a_0}\right )^2}-2\Omega_{s,0}\nonumber\\
\times\ln\left\lbrack
\Omega_{rad,0}\frac{a}{a_0}+\sqrt{\Omega_{rad,0}}\sqrt{\Omega_{s,0}
+\Omega_{rad,0}\left (\frac{a}{a_0}\right )^2}\right\rbrack\nonumber\\
+\Omega_{s,0}\ln(\Omega_{s,0}\Omega_{rad,0})=4(\Omega_{rad,0})^{3/2} H_0
t.\nonumber\\
\label{rada1}
\end{eqnarray}
At $t=0$ the universe starts from a singular state at which the
scale factor $a=0$ while the energy density $\epsilon\rightarrow +\infty$.
The scale factor increases with time while the energy density decreases with
time.

\begin{figure}[!ht]
\includegraphics[width=0.98\linewidth]{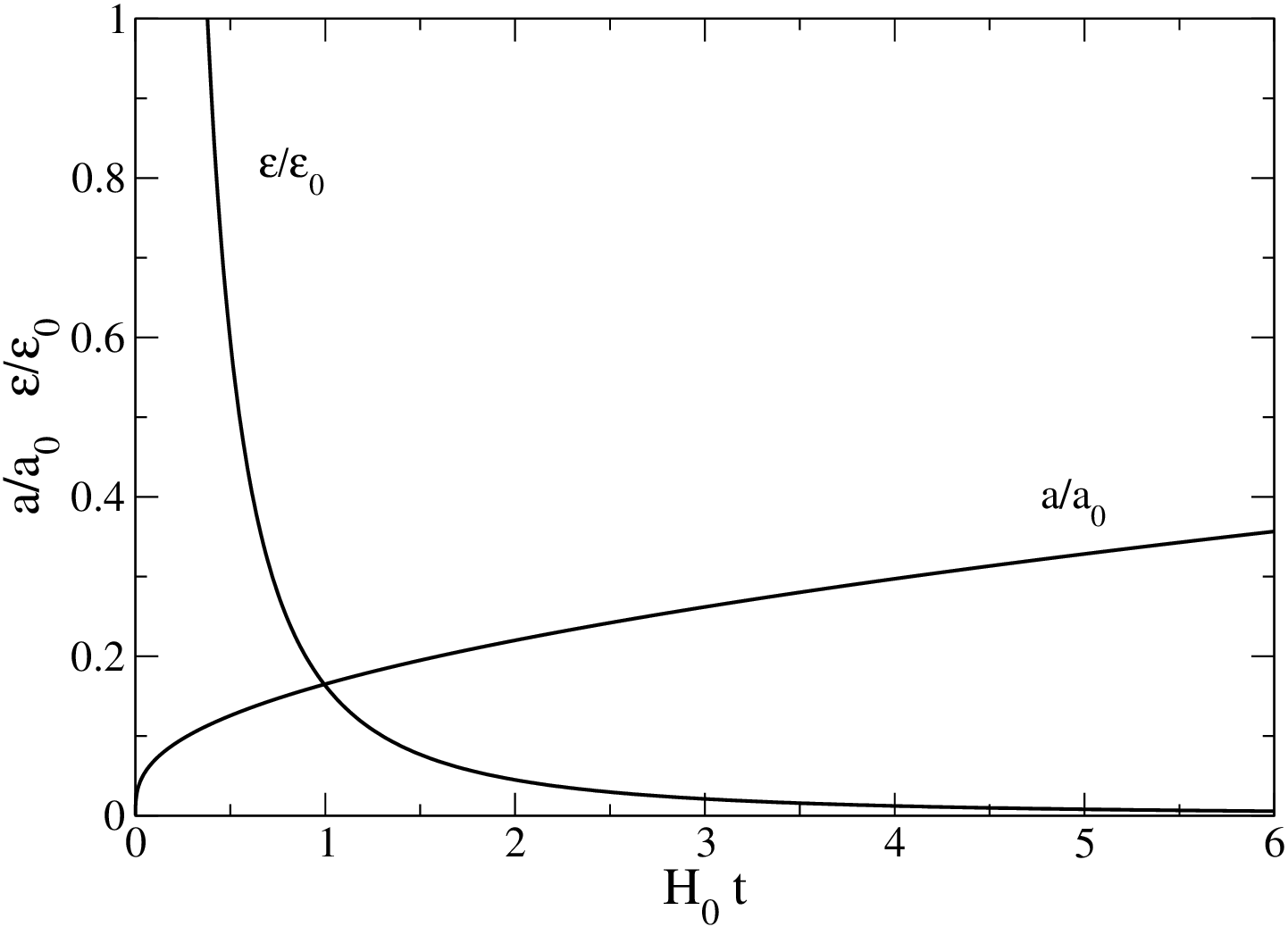}
\caption{Evolution of the scale factor and of the energy
density as a function of time. We have taken $\Omega_{s,0}=10^{-6}$ and
$\Omega_{rad,0}=8.48\, 10^{-5}$ (here and in the
following figure, we have chosen a relatively large value of the density of
stiff matter $\Omega_{s,0}$ for a better illustration of the
results). \label{radiation}}
\end{figure}

The temporal evolutions of the scale factor and of the energy density are shown
in Fig. \ref{radiation}.

\subsection{Anti-stiff matter and radiation}
\label{sec_radb}

We consider  a  universe made of anti-stiff matter and radiation. The total
energy is
\begin{equation}
\frac{\epsilon}{\epsilon_0}=-\frac{|\Omega_{s,0}|}{(a/a_0)^6}+\frac{\Omega_{rad,
0 }
}{(a/a_0)^4}.
\label{radb0}
\end{equation}
The energy density is positive for $a\ge a_i$ with
\begin{equation}
\frac{a_i}{a_0}=\left (\frac{|\Omega_{s,0}|}{\Omega_{rad,0}}\right )^{1/2}.
\label{radb0b}
\end{equation}
The energy density starts from $\epsilon=0$ at $a=a_i$, increases, reaches a
maximum at
\begin{equation}
\frac{a_*}{a_0}=\left (\frac{3|\Omega_{s,0}|}{2\Omega_{rad,0}}\right )^{1/2},
\qquad \frac{\epsilon_*}{\epsilon_0}=\frac{4 \Omega_{rad,0}^3}{27
|\Omega_{s,0}|^2},
\label{radb0c}
\end{equation}
and decreases. In the absence of dust matter and dark energy
($\Omega_{m,0}=\Omega_{\Lambda,0}=0$) the integral in Eq. (\ref{stiff3})
can be performed analytically giving
\begin{eqnarray}
2\sqrt{\Omega_{rad,0}}\frac{a}{a_0}\sqrt{-|\Omega_{s,0}|+\Omega_{rad,0}\left
(\frac{a}{a_0}\right )^2}+2|\Omega_{s,0}|\nonumber\\
\times\ln\left\lbrack
\Omega_{rad,0}\frac{a}{a_0}+\sqrt{\Omega_{rad,0}}\sqrt{-|\Omega_{s,0}|+\Omega_{
rad,0}\left (\frac{a}{a_0}\right )^2}\right\rbrack\nonumber\\
-|\Omega_{s,0}|\ln(|\Omega_{s,0}|\Omega_{rad,0})=4(\Omega_{rad,0})^{3/2} H_0
t.\nonumber\\
\label{radb1}
\end{eqnarray}
At $t=0$ the universe starts from a non-singular state at which the
scale factor $a=a_i$ and the energy density $\epsilon=0$. The scale factor
increases with time. The energy density starts from $\epsilon=0$, increases,
reaches its maximum value $\epsilon_*$ at $t=t_*$ where 
\begin{equation}
t_*=\frac{|\Omega_{s,0}|}{4\Omega_{rad,0}^{3/2}H_0}\left\lbrack
\sqrt{3}+2\ln(\sqrt{3}+1)-\ln 2\right\rbrack,
\label{radb2}
\end{equation}
and decreases.

\begin{figure}[!ht]
\includegraphics[width=0.98\linewidth]{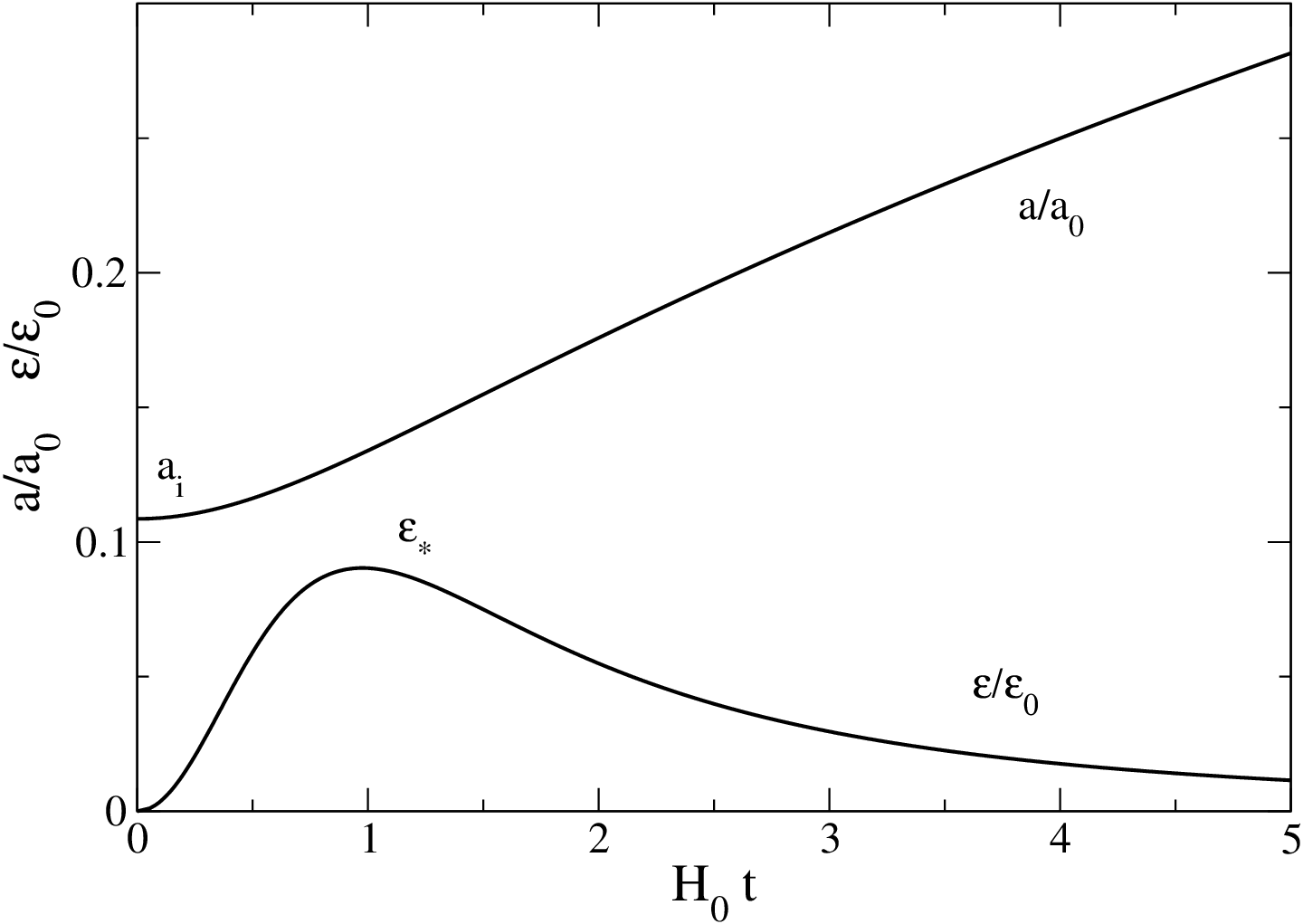}
\caption{Evolution of the scale factor and of the energy
density as a function of time. We have taken $\Omega_{s,0}=-10^{-6}$ and
$\Omega_{rad,0}=8.48\, 10^{-5}$. \label{radiationANTI}}
\end{figure}

The temporal evolutions of the scale factor and of the energy density are shown
in Fig. \ref{radiationANTI}.

\subsection{Radiation}
\label{sec_radc}

We consider  a  universe made of radiation. In the absence of stiff
matter, dust matter, and dark
energy ($\Omega_{s,0}=\Omega_{m,0}=\Omega_{\Lambda,0}=0$) we get
\begin{eqnarray}
\frac{a}{a_0}=\Omega_{rad,0}^{1/4}\sqrt{2H_0 t}, \qquad
\frac{\epsilon}{\epsilon_0}=\frac{1}{(2H_0 t)^2}.
\label{radc1}
\end{eqnarray}

\subsection{Radiation and dust matter}
\label{sec_radd}

We consider  a  universe made radiation and dust matter. The total energy is
\begin{equation}
\frac{\epsilon}{\epsilon_0}=\frac{\Omega_{rad,0}
}{(a/a_0)^4}+\frac{\Omega_{m,0}}{(a/a_0)^3}.
\label{radd0}
\end{equation}
The energy density starts from $\epsilon=+\infty$ at $a=a_i=0$ and decreases
as $a$ increases. In the absence of
stiff matter and dark energy ($\Omega_{s,0}=\Omega_{\Lambda,0}=0$) the
integral in Eq. (\ref{stiff3}) can be performed analytically
leading to
\begin{eqnarray}
H_0 t=-\frac{2}{3}\frac{1}{(\Omega_{m,0})^{1/2}}\left
(\frac{2\Omega_{rad,0}}{\Omega_{m,0}}-\frac{a}{a_0}\right
)\sqrt{\frac{\Omega_{rad,0}}{\Omega_{m,0}}+\frac{a}{a_0}}\nonumber\\
+\frac{4}{3}\frac{(\Omega_{rad,0})^{3/2}}{(\Omega_{m,0})^2}.\qquad
\label{radd1}
\end{eqnarray}
Eq. (\ref{radd1}) can also be written as
\begin{equation}
\left (\frac{a}{a_0}\right )^3-3\frac{\Omega_{rad,0}}{\Omega_{m,0}}\left
(\frac{a}{a_0}\right
)^2=\frac{9}{4}\Omega_{m,0}H_0^2t^2-6\frac{\Omega_{rad,0}^{3/2}}{\Omega_{m,0}}
H_0 t.
\label{radd2}
\end{equation}
This is a cubic equation for $a/a_0$.   At $t=0$ the universe starts from a
singular state at which the
scale factor $a=0$ while the energy density $\epsilon\rightarrow +\infty$.
The scale factor increases with time while the energy density decreases with
time.

\subsection{Radiation and dark energy}
\label{sec_ppj}

We consider a universe made of radiation and dark energy. The total energy is
\begin{equation}
\frac{\epsilon}{\epsilon_0}=\frac{\Omega_{rad,0}
}{(a/a_0)^4}+\Omega_{\Lambda,0}.
\label{ppj1}
\end{equation}
The energy density starts from $\epsilon=+\infty$ at $a=a_i=0$ and tends to
$\epsilon_{\Lambda}$ when $a\rightarrow +\infty$. In the absence of
stiff matter and dust matter ($\Omega_{s,0}=\Omega_{m,0}=0$), we get
\begin{eqnarray}
\frac{a}{a_0}=\left (\frac{\Omega_{rad,0}}{\Omega_{\Lambda,0}}\right
)^{1/4}\sinh^{1/2}\left (2\sqrt{\Omega_{\Lambda,0}}H_0 t\right ),
\label{ppj2}
\end{eqnarray}
\begin{eqnarray}
\frac{\epsilon}{\epsilon_0}=\frac{\Omega_{\Lambda,0}}{\tanh^2\left
(2\sqrt{\Omega_{\Lambda,0}}H_0 t\right )}.
\label{ppj3}
\end{eqnarray}
At $t=0$ the universe starts from a singular state at which the
scale factor $a=0$ while the energy density $\epsilon= +\infty$.
The scale factor increases with time while the energy density decreases with
time and tends to $\epsilon_{\Lambda}$ for $t\rightarrow +\infty$.

\subsection{Radiation and anti-dark energy}
\label{sec_ppk}

We consider a universe made of radiation and anti-dark energy. The total energy
is
\begin{equation}
\frac{\epsilon}{\epsilon_0}=\frac{\Omega_{rad,0}
}{(a/a_0)^4}-|\Omega_{\Lambda,0}|.
\label{ppk1}
\end{equation}
The energy density is positive for $a\le a_f$ with
\begin{equation}
\frac{a_f}{a_0}=\left (\frac{\Omega_{rad,0}}{|\Omega_{\Lambda,0}|}\right
)^{1/4}.
\label{ppk1b}
\end{equation}
The energy density starts from $\epsilon=+\infty$ at $a=a_i=0$, decreases, and
reaches $\epsilon=0$ at $a=a_f$. In the absence
of stiff matter and dust matter ($\Omega_{s,0}=\Omega_{m,0}=0$), we get
\begin{eqnarray}
\frac{a}{a_0}=\left (\frac{\Omega_{rad,0}}{|\Omega_{\Lambda,0}|}\right
)^{1/4}\sin^{1/2}\left (2\sqrt{|\Omega_{\Lambda,0}|}H_0 t\right ),
\label{ppk2}
\end{eqnarray}
\begin{eqnarray}
\frac{\epsilon}{\epsilon_0}=\frac{|\Omega_{\Lambda,0}|}{\tan^2\left
(2\sqrt{|\Omega_{\Lambda,0}|}H_0 t\right )}.
\label{ppk3}
\end{eqnarray}
At $t=0$ the universe starts from a singular state at which the
scale factor $a=0$ while the energy density $\epsilon=+\infty$.
Between $t=0$ and $t=t_1$ where 
\begin{eqnarray}
t_1=\frac{\pi}{4\sqrt{|\Omega_{\Lambda,0}|}H_0},
\label{ppk4}
\end{eqnarray}
the energy density decreases from $\epsilon=+\infty$ to $\epsilon=0$ and the
scale factor increases from $a=0$ to $a=a_f$. Between $t_1$ and $t_2=2t_1$ the
energy density increases from $\epsilon=0$ to $\epsilon=+\infty$ and the scale
factor decreases from $a=a_f$ to $a=0$. This process continues periodically
with a period $t_2$.

\section{Conclusion}

In this paper, we have obtained analytical solutions of the Friedmann equations
for a universe undergoing a primordial stiff matter era. This stiff matter era
appears in the cosmological model of Zel'dovich \cite{zeldocosmo,zeldovich} in
which the universe is made of a cold gas of baryons. It also appears in certain
models of relativistic BECs with a stiff equation of state \cite{mlbec}. In this
paper, we have studied the evolution of the homogeneous background. For the sake
of generality, we have considered a positive or a negative energy density of
the stiff matter (leading to singular or non-singular models of universe)
and a positive or a negative value of the cosmological constant (leading to
expanding or oscillating models of universe). In a future work, we shall
consider the evolution of the perturbations in these models.

\appendix 

\section{A generalization of the analytical solutions}
\label{sec_gen}

We consider a universe containing three non-interacting fluids, each
of them described by a linear equation of state $p_i=\alpha_i\epsilon_i$
with
$\alpha_1=\alpha$, $\alpha_2=(\alpha-1)/2$, and $\alpha_3=-1$ (dark energy). 
Since $\alpha_1\le 1$, we  have
$\alpha_2\le 0$. Therefore, the second fluid necessarily has a negative (or a
vanishing)
pressure. On the other hand, the two fluids are either both normal
($\alpha_1>-1$ and $\alpha_2>-1$) or both phantom-like ($\alpha_1<-1$ and
$\alpha_2<-1$). Some triplets $(\alpha_1,\alpha_2,\alpha_3)$ of physical
interest are $(1,0,-1)$, $(0,-1/2,-1)$, $(1/3,-1/3,-1)$, and $(-1,-1,-1)$. The
first case $(\alpha_1,\alpha_2,\alpha_3)=(1,0,-1)$ corresponds to a universe
made of stiff matter, dust matter, and dark energy as studied in the main part
of this
paper. The total energy density is
\begin{equation}
\frac{\epsilon}{\epsilon_0}=\frac{\Omega_{s,0}}{(a/a_0)^{3(1+\alpha)}}
+\frac{\Omega_{m,0}}{(a/a_0)^{\frac{3}{2}(1+\alpha)}}+\Omega_{\Lambda,0}.
\label{gen1}
\end{equation}
The Friedmann equation (\ref{f3}) takes the form
\begin{equation}
\int_{a_i/a_0}^{a/a_0} \frac{dx}{x
\sqrt{\frac{\Omega_{s,0}}{x^{3(1+\alpha)}}+\frac{\Omega_{m,0}}{x^{\frac{3}{2}(1+
\alpha)}} +\Omega_ { \Lambda , 0 } } } =H_0 t.
\label{gen2}
\end{equation}
With the change of variables $X=x^{(1+\alpha)/2}$, we obtain
\begin{equation}
\int_{(a_i/a_0)^{(1+\alpha)/2}}^{(a/a_0)^{(1+\alpha)/2}} \frac{dX}{X
\sqrt{\frac{\Omega_{s,0}}{X^{6}}+\frac{\Omega_{m,0}}{X^{3}}
+\Omega_ { \Lambda , 0 } } } =\frac{1+\alpha}{2} H_0 t,
\label{gen3}
\end{equation}
where we recognize the integral in Eq. (\ref{stiff3c}). As a result, the
evolution of the scale factor and of the energy density of a universe containing
three
non-interacting fluids with linear coefficient
$\alpha_1=\alpha$, $\alpha_2=(\alpha-1)/2$, and $\alpha_3=-1$ are
given by the equations of the main part of this paper with the substitutions
$a/a_0\rightarrow
(a/a_0)^{(1+\alpha)/2}$ and $H_0 t\rightarrow (1+\alpha)H_0 t/2$.

\section{General polytropic equation of state}
\label{sec_gc}

In the main part of the paper (see also \cite{mlbec}), we have considered 
a polytropic equation of state of index $n=1$ (i.e. $\gamma=2$). This is the
equation of state that appears in the model of Zel'dovich
\cite{zeldocosmo,zeldovich}. This is also the standard equation of state of a
self-interacting BEC at $T=0$ \cite{revuebec}.  More generally, we can consider
the  polytropic equation of state \cite{chandra}:
\begin{equation}
P=K\rho^{\gamma},\qquad \gamma=1+\frac{1}{n},
\label{gc1}
\end{equation} 
with arbitrary $\gamma$. For example, this equation of state can be derived
from the Gross-Pitaevskii equation 
\begin{equation}
i\hbar \frac{\partial\psi}{\partial
t}=-\frac{\hbar^2}{2m}\Delta\psi+m\left (\Phi+\frac{K\gamma}{\gamma-1}(Nm)^{
\gamma-1 } |\psi|^{2(\gamma-1)}\right )\psi,
\label{gc1b}
\end{equation} 
for a  BEC at $T=0$ with an arbitrary
nonlinearity \cite{chavaniscosmo}.

We assume that the universe is filled with a relativistic fluid at $T=0$
described by the polytropic equation of state (\ref{gc1}).  From
Eq. (\ref{gr5}), we find that the energy density is related to the rest-mass
density by \cite{mlbec,tooper2}:
\begin{equation}
\epsilon=\rho c^2+K\rho\ln(\rho/\rho_*), \qquad (\gamma=1),
\label{gc2}
\end{equation}
\begin{equation}
\epsilon=\rho c^2+\frac{K}{\gamma-1}\rho^{\gamma}=\rho c^2+nP,\qquad
(\gamma\neq 1).
\label{gc3}
\end{equation}
Combining Eq. (\ref{eos3}) and Eqs. (\ref{gc2}) and (\ref{gc3}), we obtain for
$\gamma=1$:
\begin{equation}
\epsilon=\rho_0 c^2\left (\frac{a_0}{a}\right
)^3+K\rho_0\left (\frac{a_0}{a}\right
)^3\ln\left\lbrack \frac{\rho_0}{\rho_*} \left
(\frac{a_0}{a}\right )^{3}\right\rbrack
\label{gc4}
\end{equation}
and for $\gamma\neq 1$:
\begin{equation}
\epsilon=\rho_0 c^2\left (\frac{a_0}{a}\right
)^3+\frac{K}{\gamma-1}\rho_0^{\gamma}\left
(\frac{a_0}{a}\right )^{3\gamma}.
\label{gc5}
\end{equation}
When $\gamma>1$ (i.e. $n>0$), we find that $P\sim \epsilon/n$ and $\epsilon\sim
nK\rho^{\gamma}\propto a^{-3\gamma}$ in the ``early'' universe ($a$ small) and
that $P\sim K(\epsilon/c^2)^{\gamma}$ and $\epsilon\sim \rho c^2\propto a^{-3}$
in
the ``late'' universe ($a$ large). When $\gamma<1$ (i.e. $n<0$), we find that
$P\sim K(\epsilon/c^2)^{\gamma}$ and $\epsilon\sim \rho c^2\propto a^{-3}$ in
the
``early'' universe ($a$ small) and that $P\sim \epsilon/n$ and $\epsilon\sim
nK\rho^{\gamma}\propto a^{-3\gamma}$ in the ``late'' universe ($a$ large).

Introducing appropriate notations, the Friedmann equation (\ref{f3}) can be
written as
\begin{equation}
\frac{H}{H_0}=\sqrt{\frac{\Omega'_{m,0}\ln\lbrack \rho_0
(a_0/a)^3\rbrack}{\ln\rho_0(a/a_0)^{3}}
+\frac{\Omega_{m,0}}{(a/a_0)^3}+\Omega_{\Lambda,0}}
\label{gc6}
\end{equation}
for $\gamma=1$, and as
\begin{equation}
\frac{H}{H_0}=\sqrt{\frac{\Omega_{\gamma,0}}{(a/a_0)^{3\gamma}}
+\frac{\Omega_{m,0}}{(a/a_0)^3}+\Omega_{\Lambda,0}}
\label{gc7}
\end{equation}
for $\gamma\neq 1$.

For the polytropic equation of state (\ref{gc1}), our treatment shows that the
energy density (\ref{gc5}) is the sum of two terms. An ordinary term
$\epsilon_m\propto a^{-3}$ equivalent to dust matter and a new
term $\epsilon_{\gamma}\propto a^{-3\gamma}$  depending on the polytropic index
$\gamma$. For  $\gamma=2$ (i.e. $n=1$), the new term is equivalent to stiff
matter
($\epsilon_{s}\propto a^{-6}$) as discussed in the main part of the
paper. For $\gamma=0$  (i.e. $n=-1$), the new term is equivalent to dark
energy ($\epsilon_{\Lambda}\propto 1$), as already noted in
\cite{cosmopoly1,cosmopoly2} when the pressure is constant and negative. For
$\gamma=4/3$  (i.e. $n=3$) the new term is equivalent to the radiation of an
ultra-relativistic gas ($\epsilon_{rad}\propto a^{-4}$). 

More generally, according to Eq. (\ref{gr5}),
the new term that appears in the energy equation is equal to the internal
energy
\begin{equation}
\epsilon_{new}=u(\rho)=\rho\int_0^{\rho}\frac{P(\rho')}{{\rho'}^2}\,
d\rho'.
\label{gc8}
\end{equation}
This relation clearly shows that the new term is related to pressure effects.
When  $P\simeq {\rm Cst}$, corresponding to $\gamma\rightarrow 0$ in the
polytropic model, we suggest that this new term describes dark energy.  
Our procedure may be used to obtain generalized models of dark energy by
considering different expressions of $P(\rho)$. In our approach, we
have a single ``dark fluid'' described by an equation of state $P(\rho)$
``unifying'' dark matter and dark energy.  We suggest that this dark fluid may
be in the form of relativistic  self-interacting BECs at $T=0$ although other
possibilities may be contemplated. If we assume that the energy density
$\epsilon_{new}$ of the new term is positive (in agreement with
the observations), we conclude from Eq. (\ref{gc3}) that the pressure must be
positive for $\gamma>1$ and negative for $\gamma<1$.  In particular,
for the index $\gamma\rightarrow 0$ describing dark energy, the pressure must
be negative. In the BEC interpretation, the pressure is negative when the
self-interaction is
attractive ($K<0$). This may justify equations of state with negative pressure
(such as the equation of state of dark energy) as already suggested in
\cite{chavaniscosmo}.

It may also be relevant to consider the logotropic equation of state
\cite{pud,logo}:
\begin{equation}
P=A\ln(\rho/\rho_*),
\label{gc9}
\end{equation} 
which can be viewed as the limiting form of a polytrope of index
$\gamma\rightarrow 0$ ($n\rightarrow -1$) with $K\rightarrow \infty$ such that
$A=K\gamma$ is finite \cite{logo}. In the case of BECs, using the
general relations of Ref. \cite{chavaniscosmo}, this equation of state can be
derived from a Gross-Pitaevskii equation of the form
\begin{equation}
i\hbar \frac{\partial\psi}{\partial
t}=-\frac{\hbar^2}{2m}\Delta\psi+m\left (\Phi-\frac{A}{Nm|\psi|^{2}}\right
)\psi,
\label{gc10}
\end{equation} 
i.e. with the exponent $-2$ (inverted potential) instead of $+2$ in the usual
Gross-Pitaevskii
equation. This equation can also be obtained as the limiting form of Eq.
(\ref{gc1b}) when $\gamma\rightarrow 0$ and $K\rightarrow \infty$ with
$A=K\gamma$ finite. For the logotropic equation of state (\ref{gc9}), the energy
relation (\ref{gr5}) becomes
\begin{equation}
\epsilon=\rho c^2-A\ln\left (\frac{\rho}{\rho_*}\right )-A.
\label{gc11}
\end{equation}
Combining Eq. (\ref{eos3}) and Eq. (\ref{gc11}), we obtain 
\begin{equation}
\epsilon=\rho_0 c^2\left (\frac{a_0}{a}\right
)^3-A\ln\left\lbrack \frac{\rho_0}{\rho_*}\left (\frac{a_0}{a}\right
)^3\right\rbrack -A.
\label{gc12}
\end{equation}
For $a\rightarrow +\infty$, we get $\epsilon\sim 3A\ln a$. Integrating the
Friedmann equation (\ref{f3}) we obtain the ``super de-Sitter'' asymptotic
behavior
\begin{equation}
a\propto e^{\frac{2\pi GA}{c^2}t^2}, \qquad (t\rightarrow +\infty).
\label{gc13}
\end{equation} 
This model will be studied in more detail in a specific paper.

\end{document}